# GPS Multipath Detection in the Frequency Domain

Elie Amani [(1)(2)(3)], Karim Djouani [(1) (2)], Anish Kurien [(2)]
[(1)]LISSI Laboratory, Université Paris-Est Créteil (UPEC)
Paris, France
[(2)] F'SATI/Dept. of Electrical Engineering, Tshwane University of Technology (TUT)
Pretoria, South Africa

Jean-Rémi De Boer [(3)], Willy Vigneau [(3)], Lionel Ries [(4)]
[(3)] M3 Systems
Toulouse, France
[(4)]Centre National d'Etudes Spatiales (CNES)
Toulouse, France

***Abstract*— Multipath is among the major sources of errors in precise positioning using GPS and continues to be extensively studied. Two Fast Fourier Transform (FFT)-based detectors are presented in this paper as GPS multipath detection techniques. The detectors are formulated as binary hypothesis tests under the assumption that the multipath exists for a sufficient time frame that allows its detection based on the quadrature arm of the coherent Early-minus-Late discriminator ($Q_{EmL}$) for a scalar tracking loop (STL) or on the quadrature ($Q_{EmL}$) and/or in-phase arm ($I_{EmL}$) for a vector tracking loop (VTL), using an observation window of N samples. Performance analysis of the suggested detectors is done on multipath signal data acquired from the multipath environment simulator developed by the German Aerospace Centre (DLR) as well as on multipath data from real GPS signals. Application of the detection tests to correlator outputs of scalar and vector tracking loops shows that they may be used to exclude multipath contaminated satellites from the navigation solution. These detection techniques can be extended to other Global Navigation Satellite Systems (GNSS) such as GLONASS, Galileo and Beidou.**

***Index Terms*—GPS, multipath detection, FFT-based detector, tracking error, scalar tracking loop, vector tracking loop.**

## I. INTRODUCTION

Multipath remains one of the major sources of errors in precise positioning using the Global Positioning System (GPS) and all other Global Navigation Satellite Systems (GNSS). GPS pseudorange measurements can be severely compromised by multipath signals because these signals destroy the correlation function shape used for time delay estimation. Recently, several signal processing techniques have been developed to deal with errors induced by multipath signals in a GPS receiver. Three categories of techniques can be distinguished. The first category relies on detection of multipath using statistical detection methods [1]. No mitigation is performed but the satellite whose line-of-sight (LOS) signal is absent or severely affected by secondary paths is excluded from the navigation solution. Examples of methods that fall under this category are the Early Late Phase (ELP)-based detection [2] [3], the ANOVA-based detection [4], detection based on analysis of SNR fluctuation [5], detection based on code minus carrier delta-range measurement [6]. The ANOVA method requires a multi-antenna receiver (physical or logical antenna array). With the SNR-based method, the fluctuations occur with a periodicity of 1.5 to 20 minutes for GPS satellites, which requires the multipath detection test to be set over long periods of time (10 minutes in [5]) therefore delaying multipath detection. The second category of techniques modifies the receiver tracking loop to make it resistant to multipath signals. Under this category fall the Narrow Correlator [7], the Edge Correlator [8], the Strobe Correlator [8], the High Resolution Correlator (HRC) [9], the Gated Correlator [10], the Multipath Elimination Technology (MET) [11], and the A-Posteriori Multipath Estimation (APME) Technique [12]. The third category relies on joint detection and estimation of the line-of-sight (LOS) and/or multipath signal parameters (amplitude, delay and phase) using statistical estimation methods [13]. It includes the Multipath Estimating Delay Lock Loop (MEDLL) [14], the Modified RAKE Delay Lock Loop (MRDLL) [15], the Multipath Mitigation Technology (MMT) [16], the Vision Correlator (VC) [17], the Fast Iterative Maximum-Likelihood Algorithm (FIMLA) [18], Deconvolution Approaches [19] [20], and Frequency Domain Processing [21] [22].

The techniques suggested by this paper are multipath detection techniques. They are suitable for a mono or multi antenna receiver and the time delay before multipath detection is significantly small (between 30 ms and 1 sec) compared to the technique in [5]. No multipath mitigation is performed. Instead, these detection techniques may be used to exclude multipath affected satellites from calculation of the position, velocity, and time (PVT) solution. In fact, multipath estimation techniques, especially those based on frequency domain processing, may present a high computational burden compared to multipath detection techniques. And the fact that many satellite constellations can now be jointly used to obtain the PVT solution in one GNSS receiver, excluding multipath contaminated satellites instead of spending computational resources to mitigate multipath can sometimes be a more handy approach. The detectors avoid the additional computational burden of multipath parameters estimation but still incur the computational cost of Fast Fourier Transform (FFT) calculation. The detection tests are proposed for both scalar and vector tracking loops (STL and VTL). The detectors indications can also be used to switch between STL and VTL tracking modes. The detectors are based on correlator outputs, meaning that the test metrics are defined using correlator outputs. For a scalar tracking loop (STL) utilizing a coherent Early-minus-Late (EmL) discriminator, the EmL correlation output is directly used in the detectors metrics. Indeed, for a STL, depending on the relative phase of the multipath signal, the presence of multipath increases the signal power on the quadrature arm of the EmL correlation point ($Q_{EmL}$) compared to the power that is usually observed in the absence of multipath. This increase in signal power is used to detect multipath with the specification of a proper threshold. For a vector tracking loop (VTL) using the coherent EmL discriminator, both the in-phase and quadrature arms of the EmL correlation outputs are utilized in the detectors metrics. In fact, for a VTL, the increase in signal power due to the presence of multipath may occur on the in-phase arm ($I_{EmL}$) and/or on the quadrature arm ($Q_{EmL}$) of the EmL correlation output depending on the delay and phase of the multipath signal. The tracking scheme used in this paper is a vector delay frequency locked loop (VDFLL) assisted by a scalar phase locked loop (PLL), with the possibility to switch to a scalar delay locked loop (DLL) and a frequency-assisted phase locked loop (FPLL).

The rest of the paper is organized as follows. Section II presents the multipath model that was used to theoretically assess the performance of the proposed detection techniques before applying them to DLR multipath data and real GPS multipath signal data. Section III explains the link between correlator outputs and multipath detection. In section IV, the detection approach used for each detector is described in detail. Experimental results are provided in section V and section VI gives concluding remarks.

## II. Multipath Model and correlator outputs

If a specular multipath model with a finite number of multipath signals is considered, the signal entering the code and phase loop, neglecting the low rate data, is expressed as:

$$x(t) = A_0 C(t-\tau_0)\cos(\omega t - \varphi_0) + \sum_{l=1}^{L} A_l C(t-\tau_l)\cos(\omega t - \varphi_l) + w(t) \quad (1)$$

where $L$ is the number of multipath signals, $A_0$ and $A_l$ are the line-of-sight (LOS) and $l^{th}$ multipath amplitudes respectively, $C(t)$ is the spreading code, $\tau_0$, $\tau_l$, $\varphi_0$, $\varphi_l$ are the time and phase delays induced by the transmission from satellite to receiver for the LOS and $l^{th}$ multipath signals respectively, and $\omega$ is the nominal GPS L1, L2 or L5 pulsation and $w(t)$ is the zero-mean additive white Gaussian noise with variance $\sigma^2$. The signal at the output of the prompt correlator at an instant of time is therefore:

$$U_P = A_0 R(\Delta\tau)\exp(j\Delta\varphi) + \sum_{l=1}^{L} A_l R(\Delta\tau - \delta_l)\exp[j(\Delta\varphi - \theta_l)] + w_P \quad (2)$$

where $R$ is the correlation function, $\Delta\tau$ is the error between the LOS signal delay and the estimated code replica delay, $\Delta\varphi$ is the error between the LOS carrier phase and the estimated carrier replica phase, $\delta_l$ is the delay of the $l^{th}$ multipath with respect to the LOS, $\theta_l$ is the phase shift of the $l^{th}$ multipath with respect to the LOS, and $w_P$ is the post-correlation noise. For the rest of the paper a one-path specular multipath model is used. In fact, it can be assumed that all reflected signals have the same frequency and combine into a single effective multipath. In that case, the signal entering the code and phase loops can be expressed as:

$$x(t) = A_0 C(t-\tau_0)\cos(\omega t - \varphi_0) + A_M C(t-\tau_M)\cos(\omega t - \varphi_M) + w(t) \quad (3)$$

where the subscript $M$ refers to the multipath resulting from a vector sum of all multipath signals in presence. The multipaths with a different frequency would contribute to an increase in noise power. Thus, the in-phase and quadrature outputs of the prompt correlator in the presence of a specular multipath, at an instant of time, are:

$$I_P = A_0 R(\Delta\tau)\cos(\Delta\varphi) + A_M R(\Delta\tau - \delta_M)\cos(\Delta\varphi - \theta_M) + w_{I,P} \quad (4a)$$

$$Q_P = A_0 R(\Delta\tau)\sin(\Delta\varphi) + A_M R(\Delta\tau - \delta_M)\sin(\Delta\varphi - \theta_M) + w_{Q,P} \quad (4b)$$

Similarly, the Early and Late in-phase and quadrature correlator outputs are given by:

$$I_E = A_0 R(\Delta\tau + d)\cos(\Delta\varphi) + A_M R(\Delta\tau - \delta_M + d)\cos(\Delta\varphi - \theta_M) + w_{I,E} \quad (5a)$$

$$Q_E = A_0 R(\Delta\tau + d)\sin(\Delta\varphi) + A_M R(\Delta\tau - \delta_M + d)\sin(\Delta\varphi - \theta_M) + w_{Q,E} \quad (5b)$$

$$I_L = A_0 R(\Delta\tau - d)\cos(\Delta\varphi) + A_M R(\Delta\tau - \delta_M - d)\cos(\Delta\varphi - \theta_M) + w_{I,L} \quad (6a)$$

$$Q_L = A_0 R(\Delta\tau - d)\sin(\Delta\varphi) + A_M R(\Delta\tau - \delta_M - d)\sin(\Delta\varphi - \theta_M) + w_{Q,L} \quad (6b)$$

where $d$ is half the Early-Late chip spacing $\partial = 2d$ ($0 < \partial \le 1$). The EmL in-phase and quadrature outputs are obtained by subtracting Eq. (6) from Eq. (5):

$$I_{EmL} = A_0\left[R(\Delta\tau + d) - R(\Delta\tau - d)\right]\cos(\Delta\varphi) + A_M\left[R(\Delta\tau - \delta_M + d) - R(\Delta\tau - \delta_M - d)\right]\cos(\Delta\varphi - \theta_M) + w_{I,EmL} \quad (7a)$$

$$Q_{EmL} = A_0\left[R(\Delta\tau + d) - R(\Delta\tau - d)\right]\sin(\Delta\varphi) + A_M\left[R(\Delta\tau - \delta_M + d) - R(\Delta\tau - \delta_M - d)\right]\sin(\Delta\varphi - \theta_M) + w_{Q,EmL} \quad (7b)$$

For a locked VTL, it can be assumed that $\Delta\tau \approx 0$, which implies that $R(\Delta\tau + d) - R(\Delta\tau - d) \approx 0$ as well. The EmL in-phase and quadrature outputs in this case are:

$$I_{EmL} = A_M\left[R(-\delta_M + d) - R(-\delta_M - d)\right]\cos(\Delta\varphi - \theta_M) + w_{I,EmL} \quad (8a)$$

$$Q_{EmL} = A_M\left[R(-\delta_M + d) - R(-\delta_M - d)\right]\sin(\Delta\varphi - \theta_M) + w_{Q,EmL} \quad (8b)$$

While for a STL, the quadrature arm of the coherent EmL discriminator output (7b) is used in the detectors metrics that are described in section IV; for a VTL, the modulus of the complex output of the coherent EmL discriminator is used, i.e. the in-phase and quadrature EmL outputs (8a and 8b) are taken into account.

$$\begin{aligned} EmL &= I_{EmL} + jQ_{EmL} = (I_E - I_L) + j(Q_E - Q_L) \\ &= A_M\left[R(-\delta_M + d) - R(-\delta_M - d)\right]\exp\left[j(\Delta\varphi - \theta_M)\right] + w_{EmL} \end{aligned} \quad (9a)$$

$$|EmL| = \sqrt{\left(I_{EmL}{}^2 + Q_{EmL}{}^2\right)} \quad (9b)$$

## III. Correlator Outputs And Multipath detection

A careful analysis of GPS correlator outputs in the absence and in the presence of multipath can lead to the design of multipath detection techniques.

In the absence of multipath, the in-phase prompt correlator output carries the LOS signal power. In the presence of a multipath (MP) signal, the prompt correlator output is composed of the sum of the LOS and MP signals and the STL locking point is adjusted to this sum. As the tracking loop constantly seeks to bring the quadrature prompt power to zero, the quadrature prompt output will have part of the LOS + MP signal power only for a short transient time following MP arrival then will get back to zero, unless the MP signal is in phase or opposition of phase with the LOS signal.

The situation in the presence of multipath is different however for early and late correlator outputs and consequently for in-phase EmL and quadrature EmL outputs as can be observed in Fig. 1 for the STL and in Fig. 2 for the VTL. For the STL, in the presence of multipath, the signal energy in the quadrature arm of the EmL correlator output during transient and steady-state times following multipath arrival is significantly higher than in the absence of multipath, unless the MP signal is in phase or opposition of phase with the LOS signal. In the absence of multipath, only noise is observed on the quadrature EmL output.

Figure 1 represents the $Q_{EmL}$ output with a reduced number of oscillations in order to show the oscillatory nature of the $Q_{EmL}$ output within the envelope. Instead of 1575 cycles per C/A code chip for GPS L1, 25 cycles per C/A code chip are considered. The figure shows that the $Q_{EmL}$ output is zero for some multipath phase values even in the presence of multipath. More specifically, the $Q_{EmL}$ output is zero when the multipath signal is in phase ($\theta_M = 0^\circ + k360^\circ$) or opposition of phase ($\theta_M = 180^\circ + k360^\circ$) with the LOS signal, with $k$ being an integer. Except for those phase values, the $Q_{EmL}$ output in the presence of multipath oscillates along the different multipath delay values between a maximum and a minimum which depend on the multipath to LOS amplitude ratio $\alpha$ and on the Early-Late correlator chip spacing $\partial$. These multipath phase values ($\theta_M = 0^\circ + k360^\circ$ and $\theta_M = 180^\circ + k360^\circ$) correspond to multipath delay values $\delta_M = nR_c / f_{L1}$ and $\delta_M = (n + 0.5)R_c / f_{L1}$, with $n$ an integer, if multipath phase is related to multipath delay using $\theta_M = 2\pi f_{L1}\delta_M / R_C$, i.e. if it is assumed that the multipath phase is only due to the differential path delay. In general, at the moment of reflection or diffraction, the multipath signal undergoes a relative phase $\theta_M$ that can be modelled using the differential path delay and reflector and antenna parameters [23] or else assumed random [24]. For the VTL, both the $I_{EmL}$ and $Q_{EmL}$ outputs increase in signal power in the presence of multipath as shown in Fig. 2 and when $Q_{EmL}$ is at zero, $I_{EmL}$ is not and vice versa. This means that for all multipath phase or delay values, the absolute value $|EmL| = \sqrt{I_{EmL}^2 + Q_{EmL}^2}$ increases in amplitude in the presence of multipath.

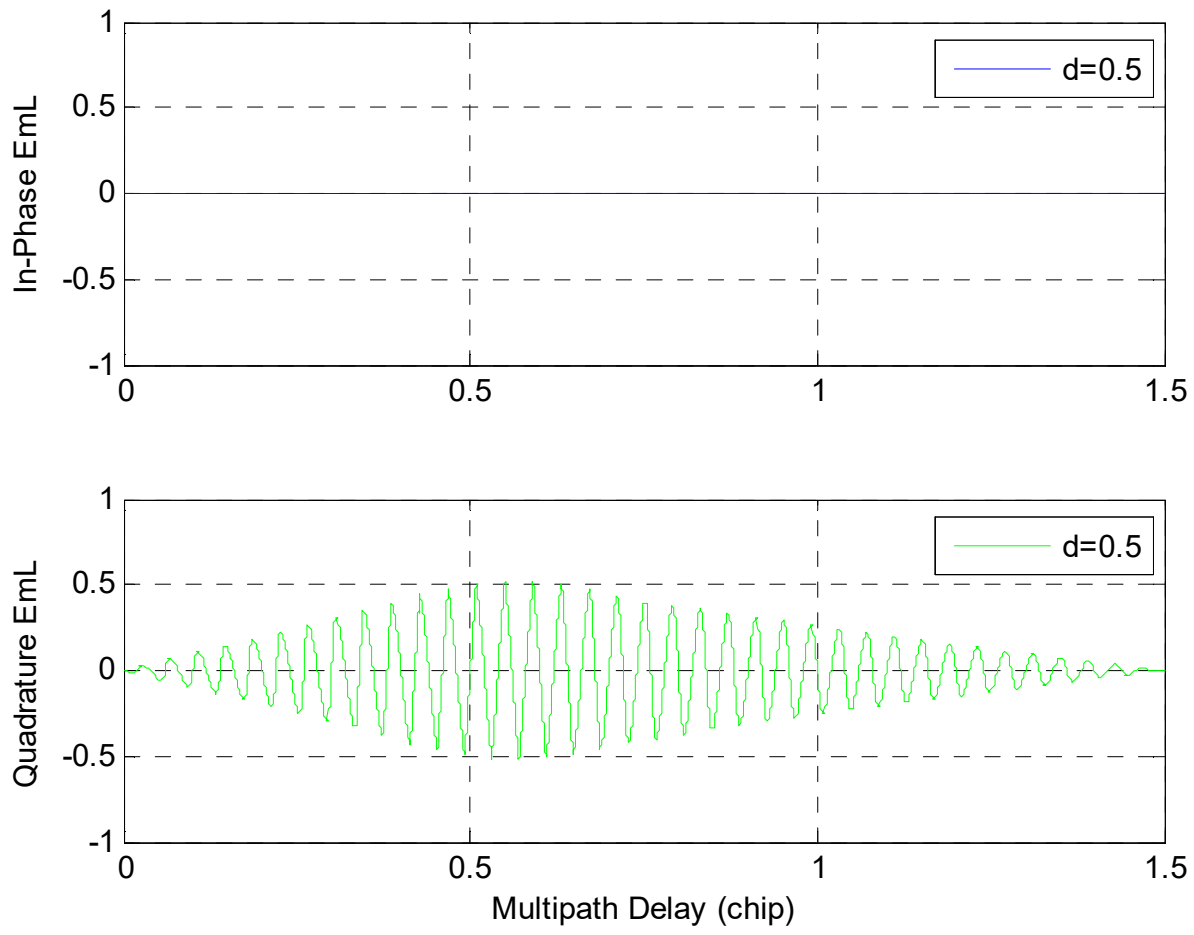

Fig. 1. $I_{EmL}$ and $Q_{EmL}$ amplitudes vs. Multipath delay for STL in lock (reduced oscillations).

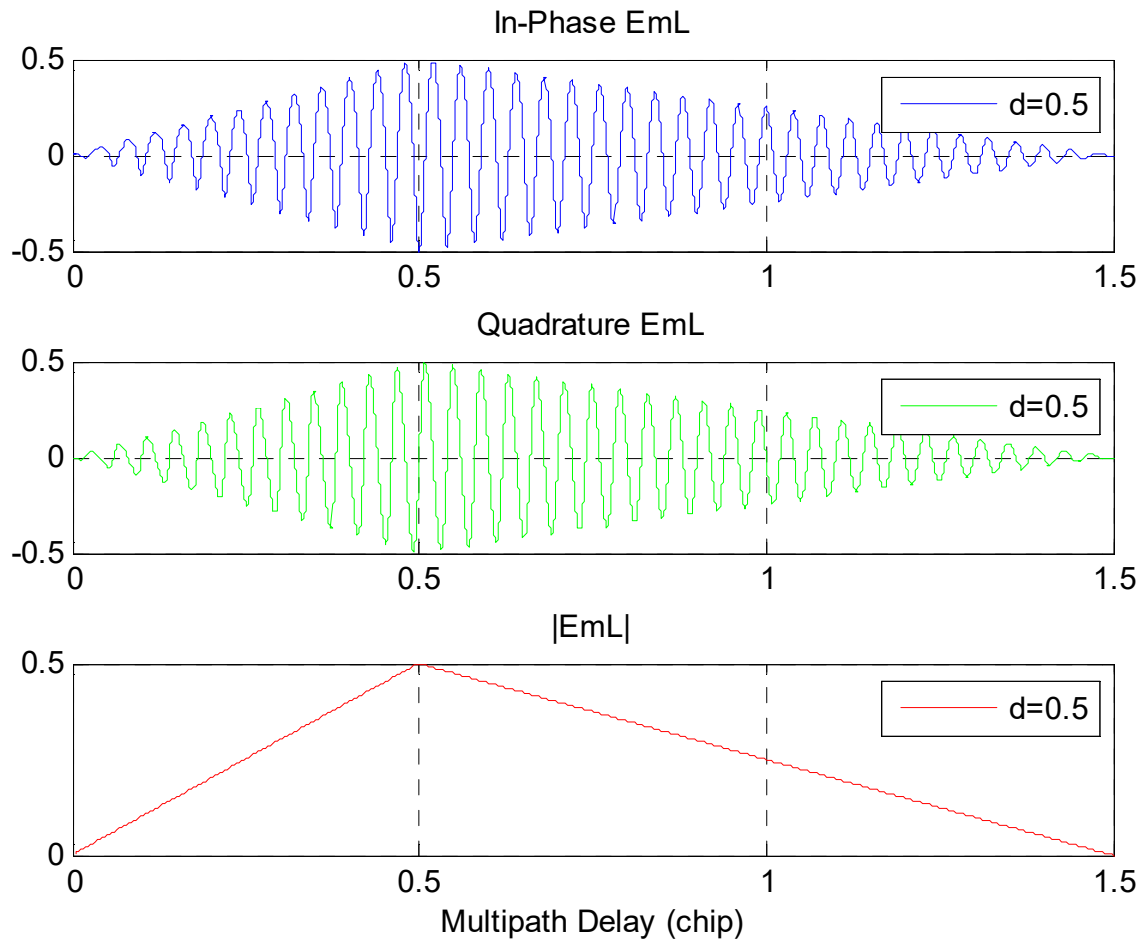

Fig. 2. $I_{EmL}$, $Q_{EmL}$ and |EmL| amplitudes vs. Multipath delay for VTL in lock (reduced oscillations).

## IV. FFT Test for Multipath Detection

### A. Detection Test on Scalar Tracking Loop (STL)

A binary hypothesis test can be defined as follows for an observation window of $N$ samples and of initial index $n_0$:

$H_0$: $Q_{EmL}(n) = w_{Q,EmL}$
$H_1$: $Q_{EmL}(n) = s(n) + w_{Q,EmL}$, with $n \in \{1, \ldots, N\} + n_0$

In other terms, under hypothesis $H_0$, no signal is present on the $Q_{EmL}$ output, only noise is observed which implies that no multipath signal is present; whereas under hypothesis $H_1$ a signal s(n) plus noise are observed, which implies the presence of multipath. The $Q_{EmL}$ post-correlation noise term $w_{Q,EmL}$ is zero-mean white Gaussian with variance $\sigma^2$.The problem is to decide for $H_1$ or $H_0$ from $Q_{EmL}$, i.e. to decide whether s(n) is present or not. Two frequency-domain MP detectors are suggested.

1) STL-MP Detector I

First, the standard Fast Fourier Transform (FFT) detector [1] [25], used in the sonar and radar systems, which detects the presence of a signal in the frequency domain is applied to GPS MP detection. This detector computes the periodogram

$$\Omega(m) = \frac{1}{N}\left|Q_{EmL}(m)\right|^2 \quad where \quad Q_{EmL}(m) = \sum_{n=n_0}^{n_0+N-1} Q_{EmL}(n) e^{-j\frac{2\pi mn}{N}} \tag{10}$$

$Q_{EmL}(m)$ is the discrete Fourier transform of $Q_{EmL}(n)$ and is implemented via an $N$-point FFT algorithm. The detector subsequently chooses the largest value of $\Omega(m)$, normalizes it with the maximum likelihood estimate (MLE) of noise variance $\hat{\sigma}^2$, and then compares the result against a threshold $\eta$. The detection test to decide for $H_0$ or $H_1$ is therefore formulated as:

$$\frac{\max(\Omega(m))}{\hat{\sigma}^2} \underset{H_1}{>} \eta \quad or \quad \frac{\max(\Omega(m))}{\hat{\sigma}^2} \underset{H_0}{<} \eta \tag{11}$$

where

$$\hat{\sigma}^2 = \frac{1}{N}\sum_{n=1}^{N}\left(Q_{EmL}(n+n_0) - \overline{Q}\right)^2 \quad and \quad \overline{Q} = \frac{1}{N}\sum_{n=1}^{N} Q_{EmL}(n+n_0)$$

Detector I is optimum if s(n) is a sinusoid. This sinusoid may have unknown amplitude, phase and frequency, but optimality requires the frequency to be a bin frequency, i.e. to be equal to $\frac{2\pi m}{N}$, with $m$ being an integer [25]. Optimum detection entails that for a given probability of false alarm (PFA), the detector gives the maximum probability of detection (PD).

The detection threshold is

$$\eta = \exp\left\{\frac{\left[cdf^{-1}(1 - PFA/2)\right]^2}{N}\right\} - 1 \tag{12}$$

with $cdf$ being the cumulative distribution function of the standard normal distribution, $cdf^{-1}$ the inverse cumulative distribution function, and PFA the probability of false alarm. The values of PFA and $N$ have a great impact on the performance of Detector I, and there is a trade-off between obtaining a high detection capability or a low false alarm rate. Increasing the value of $N$ and/or PFA improves the detection capability of Detector I in theory. However, in practice, increasing the value of $N$ delays the instant, after system initialisation, when the multipath is detected, and increases the complexity of the $N$-point FFT computation. On the other hand, increasing PFA may result in many false detections and this may not be beneficial if the objective is to exclude the multipath contaminated satellites from the navigation solution. One might wind up with fewer satellites than needed to compute the PVT solution. If $N$ is assumed sufficiently large ($N > 32$), the probability of detection (PD) is given by [1]:

$$PD = 2 - cdf\left(cdf^{-1}\left(1 - \frac{PFA}{2}\right) - \sqrt{SNR}\right) - cdf\left(cdf^{-1}\left(1 - \frac{PFA}{2}\right) + \sqrt{SNR}\right) \tag{13}$$

where $SNR$ is the post-correlation signal to noise ratio of the $Q_{EmL}$ output and is given by $SNR = \frac{NA^2}{2\sigma^2}$. This PD depends on the chosen PFA, on the amplitude $A$ of the signal s(n) on the $Q_{EmL}$ output, and on the noise power on that output. This noise power is given by $\sigma^2 = \sigma_n{}^2 K(1-r)$ [7] where $\sigma_n{}^2 = N_0 f_S$ is the thermal noise power ($N_0$ and $f_S$ are noise spectral density and baseband sampling frequency respectively), $K$ is the number of correlation points, and $r = 1 - 2d$ is the level of correlation between the Early and Late outputs. Taking, in Eq. (7b), $A_M = \alpha A_0$, the amplitude A=s(n) is given by $A = A_0\left\{\begin{matrix}[R(\Delta\tau + d) - R(\Delta\tau - d)]\sin(\Delta\varphi) + \\ \alpha[R(\Delta\tau - \delta_M + d) - R(\Delta\tau - \delta_M - d)]\sin(\Delta\varphi - \theta_M)\end{matrix}\right\}$ where $A_0 = \sqrt{C}K\,\mathrm{sinc}(\pi\Delta f T)$ with $C$ the LOS power before correlation, $T$ the coherent integration time ($K = Tf_S$) and $\Delta f$ the error between the LOS carrier frequency and the estimated carrier replica frequency. The post-correlation $SNR$ at the $Q_{EmL}$ output has thus the following expression linking it to the $C/N_0$:

$$SNR = \frac{NA^2}{2\sigma^2} = \frac{C}{N_0}\frac{NT}{2}\frac{\mathrm{sinc}^2(\pi\Delta f T)}{2d} \times \left\{\begin{matrix}[R(\Delta\tau + d) - R(\Delta\tau - d)]\sin(\Delta\varphi) + \\ \alpha[R(\Delta\tau - \delta_M + d) - R(\Delta\tau - \delta_M - d)]\sin(\Delta\varphi - \theta_M)\end{matrix}\right\}^2 \tag{14}$$

Figure 3 illustrates the probability of detection (PD) given different PFA values for different values of $SNR$. It can be observed that a high $SNR$ value increases PD and theoretically the performance of Detector I.

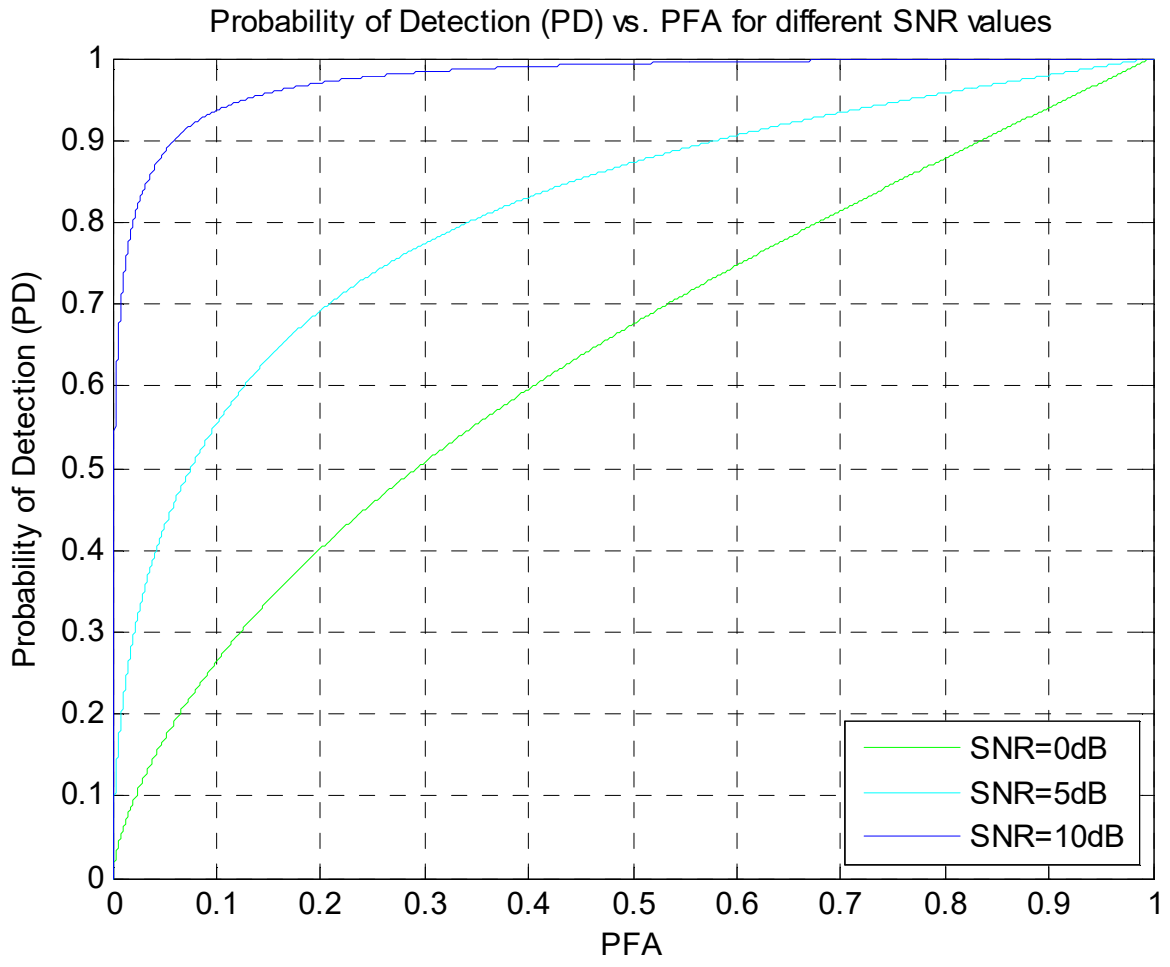

Fig. 3. Probability of multipath detection versus PFA for different SNRs (Detector I)

2) STL-MP Detector II

The second MP detector computes an FFT on blocs of 1024 samples at the frequency of correlators. In practice, this 1024-point FFT is a very complex calculation, but for the simplicity of algorithm description, it is chosen. The obtained frequency resolution is ~1 Hz for a coherent integration time of 1 ms. A power spectral density (PSD) estimator is then derived from the FFT bloc output for $|f| \leq 500$ Hz. A periodogram can be used as a PSD estimator. This PSD estimator allows the search for s(n) signal frequency in the zone of interest, by measuring the signal power using a sliding window of length 3. More specifically, this PSD estimator makes it possible:

- To estimate the noise power by summing samples of the spectral density for $|f| > 200$ Hz.
- To estimate the power of the signal by summing samples of the spectral density for $|f| < 200$ Hz.
- To estimate a signal to noise power ratio and compare it against a threshold $\eta$.

Detector II algorithm can therefore be summarized as follows:

- Compute 1024-point FFT on blocs of 1024 samples at the frequency of correlators

$$Q_{EmL}(m) = \sum_{n=n_0}^{n_0+N-1} Q_{EmL}(n) e^{-j\frac{2\pi mn}{N}} \quad where \quad N = 1024 \tag{15}$$

- Derive a periodogram (PSD estimator) for $|f| \leq 500$ Hz. Letting $m$ denote the frequency f,

$$\Omega(m) = \frac{1}{N} \left| Q_{EmL}(m) \right|^2 \tag{16}$$

- Estimate the noise power

$$NP = \sum_{m=-500}^{-200} \Omega(m) + \sum_{m=200}^{500} \Omega(m) \tag{17}$$

- Estimate the signal power

$$SP = \sum_{m=-199}^{199} \Omega(m) \tag{18}$$

- Estimate a signal to noise power ratio

$$SNPR = \frac{SP}{NP} \tag{19}$$

- Decide for $H_0$ or $H_1$ using the following test:

$$10\log(SNPR) \underset{H_1}{>} \eta \quad or \quad 10\log(SNPR) \underset{H_0}{<} \eta \tag{20}$$

where $\eta = -\ln(PFA)$ is the detection threshold. The probability of detection (PD) is given by [1]:

$$PD = 1 - CDF\left(2\ln\left(\frac{N/2-1}{PFA}\right)\right) \tag{21}$$

where $CDF$ is the cumulative distribution function of the non-central chi-squared distribution with 2 degrees of freedom and non-centrality parameter $\lambda = SNR = \frac{NA^2}{2\sigma^2}$. $SNR$ is the post-correlation signal-to-noise ratio. Figure 4 displays Detector II's probability of detection (PD) given different PFA values for different values of $SNR$.

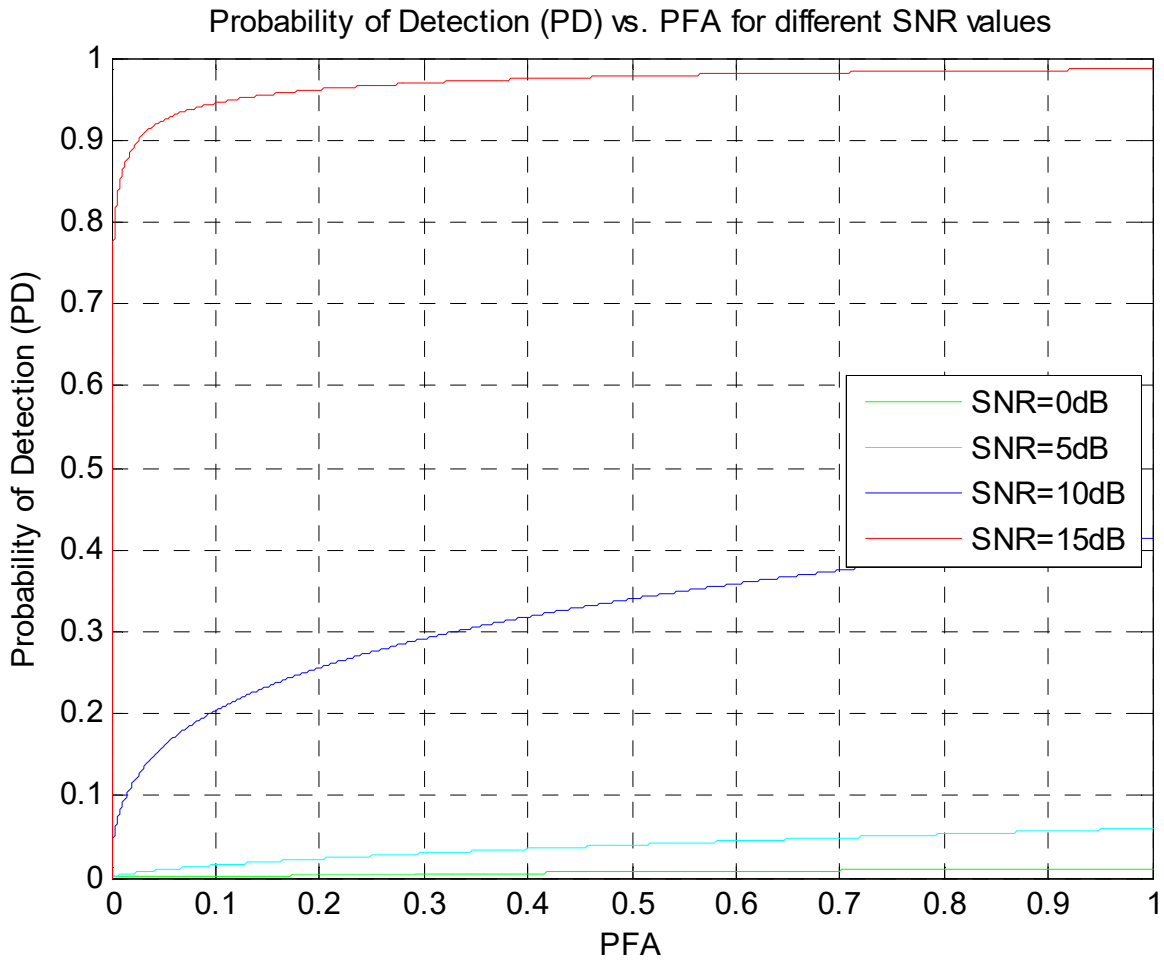

Fig. 4. Probability of multipath detection versus PFA for different SNRs (Detector II)

### *B. Detection Test on Vector Tracking Loop (VTL)*

A binary hypothesis test can be defined as follows for an observation window of $N$ samples and of initial index $n_0$:

$H_0$: EmL (n) = $w_{EmL}$

$H_1$: EmL (n) = s(n) + $w_{EmL}$, with n ϵ {1, …, $N$ } + $n_0$

The detection test to decide for $H_0$ or $H_1$ is formulated as:

$$10\log(SNPR) \underset{H_1}{>} \eta \quad or \quad 10\log(SNPR) \underset{H_0}{<} \eta \tag{22}$$

where $\eta = -\ln(PFA)$ is the detection threshold, and $SNPR$ is obtained as in the test for STL-MP Detector II except that the periodogram is calculated for $|EmL(n)| = \sqrt{(I_{EmL}(n)^2 + Q_{EmL}(n)^2)}$ and not for $Q_{EmL}(n)$. In other words,

$$EmL(m) = \sum_{n=n_0}^{n_0+N-1} |EmL(n)| e^{-j\frac{2\pi mn}{N}} \quad where \quad N = 1024 \qquad (23)$$

And,

$$\Omega(m) = \frac{1}{N} |EmL(m)|^2 \qquad (24)$$

## V. RESULTS

### A. DLR Multipath Environment Simulator Data

The DLR multipath environment simulator has been developed by The German Aerospace Center (German: Deutsches Zentrum für Luft- und Raumfahrt), abbreviated DLR. This DLR simulator generates signals that have the characteristics of GNSS signals collected in semi-urban and urban environments for a fixed-point, a moving person or a moving vehicle. These synthesized signals have the specific purpose of providing a tool for multipath study and analysis. They are manipulated then fed to the input of the code and carrier tracking loops in Matlab. These synthesized signals with the variety of multipath environment scenarios that they represent are used to evaluate the validity of the proposed detection techniques.

Some scenarios involve a vehicle moving parallel to the buildings in an urban environment and a vehicle moving in an unconstrained (open sky) environment. These are analysed in Figures 5 to 14. Figures 5, 7, 9, 11 and 13 depict the parameters (amplitude, phase and delay) of the different signal paths (LOS and/or multipath) for the scenarios and channels under study. Amplitude values are normalized, with the LOS amplitude having a maximum value of 1 (0 dB). The LOS amplitude is represented in blue in all figures depicting parameters throughout the paper. The LOS phase and delay parameters are represented either in blue or cyan colour. Both Detectors I and II perform successfully as illustrated by figures 6, 8, 10, 12 and 14. Figure 5 through the displayed amplitude, phase and delay parameters shows the presence of many multipath signals apart from the LOS signal for a channel (labelled Channel 1) tracking a satellite in STL mode for a duration of 60 seconds. Figure 6 shows the detection tests results for that channel. For simplicity of performance comparison between the two detectors, a sliding window of N=1024 samples and PFA values of $10^{-4}$ (Detector I) and $10^{-2}$ (Detector II) are used to calculate the test metric values and set the detection thresholds. The test metric values go above the defined thresholds meaning that the presence of multipath is detected for both Detectors I and II. The signal power observed on the $Q_{Eml}$ arm whose magnitude is higher than the signal power that would be observed in the absence of multipath (see Fig. 12 for comparison) also confirms the presence of multipath. In fact, in the absence of multipath, only noise is observed on the $Q_{EmL}$ arm. It is important to mention that for faster MP detection (after system initialisation) and lower computation complexity, smaller values of N such as N = 32, 64, 128, 256 can be used for Detector I, some without compromising the detection capability in low SNR scenarios. N=1024 is used with Detector II for the sake of obtaining a frequency resolution of ~1 Hz, but again smaller values of N may be used with some risk of reduced detector performance.

Two other scenarios from the DLR multipath environment simulator are defined to allow multipath to appear first around the 30$^{th}$ tracking second (see Figures 7 and 8 for STL - Channel 7 and Figures 13 and 14 for VTL - Channel 1) and then around the 20$^{th}$ tracking second (see Figures 9 and 10 for STL - Channel 9). The detectors perform as expected. The test metric values start below threshold for the 30 (respectively 20) tracking seconds where only the LOS signal is present then go above threshold around the 30$^{th}$ (respectively 20$^{th}$) second and the rest of the time. This is in accordance with the signal power increase on the $Q_{EmL}$ arm (STL case) and the increase in the absolute value of $EmL$ (VTL case) which also occur around the 30$^{th}$ (respectively 20$^{th}$) tracking second. Figures 11 and 12 are a case of a moving car in an unconstrained environment (open sky) and the tracking mode is STL. The test metric values remain below the thresholds for both Detectors I and II, meaning that no multipath is detected as expected. The signal power on the $Q_{EmL}$ arm remains minimal (made of noise only), which confirms the absence of multipath.

Figures 15 through 20 illustrate detection test results for a pedestrian moving in an urban environment. Figures 15, 16 and 17 show the case of a channel that is severely affected by multipath. At time intervals where the LOS signal power is so weak compared to multipath signals power or where the LOS is absent, the detection tests in STL tracking mode fail to detect the presence of multipath as attested by observing Fig. 15 and Fig. 16. In fact the STL loses track, as can be confirmed by the delay and phase lock indicators in Fig. 21, which renders the detection tests useless. Figure 17 shows that the VTL manages to keep tracking the signal on the same severely affected channel in Fig. 15 even at the time intervals where the LOS signal is absent or has very weak power. Detector II test on the VTL even manages to succeed in detecting multipath where it couldn't with the STL tracking mode. This result is consistent with the VTL tracking robustness in multipath environments compared to the STL. The same phenomenon is observed on Fig. 18, 19 and 20. Again a severely affected channel as shown by the parameters displayed in Fig. 18 is studied. The detection tests performed on the STL fail to detect multipath as shown in Fig. 19. The channel is severely affected to the point of having the STL diverge, and even lose lock, making the detection tests unusable. Once more the VTL manages to keep tracking the signal on the channel and to detect multipath as illustrated in Fig. 20.

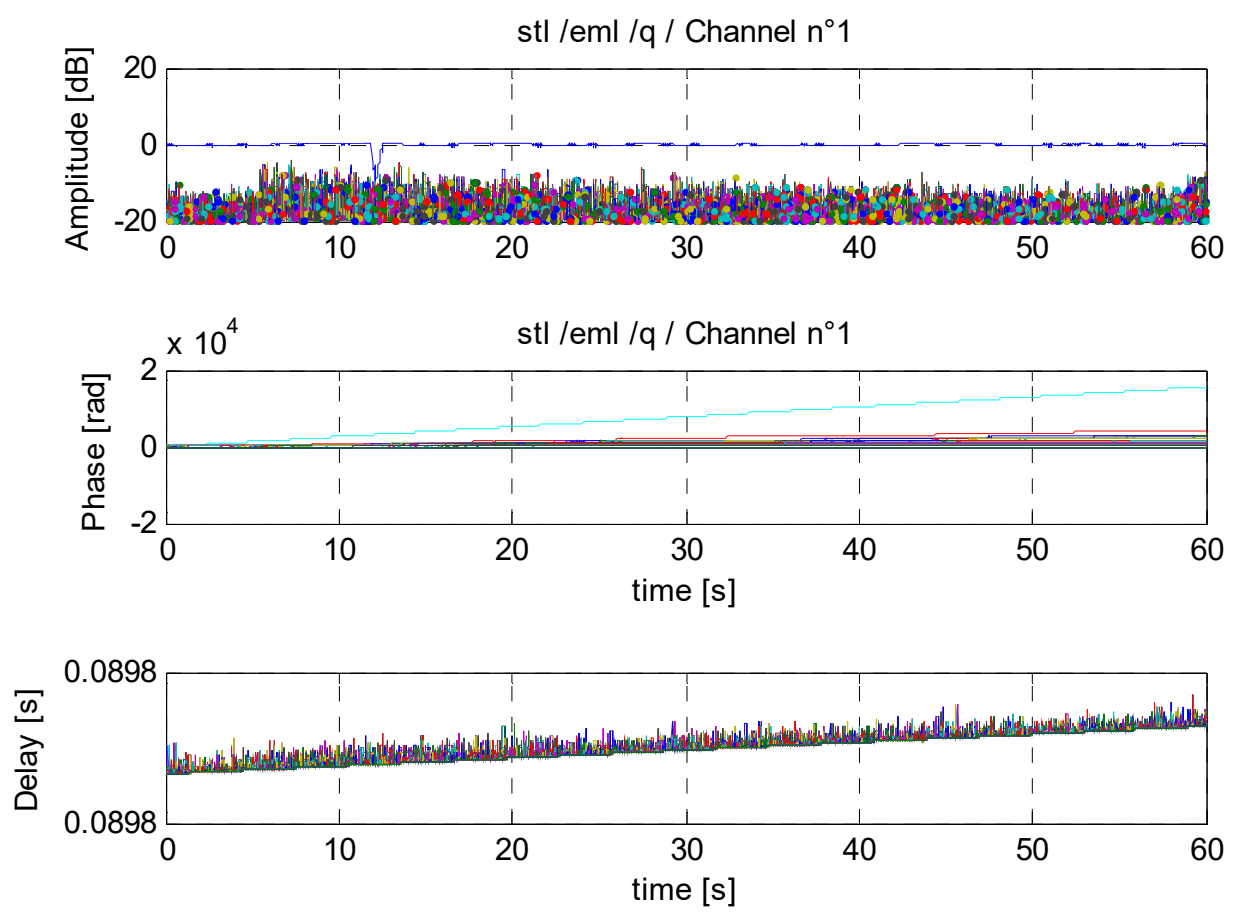

Fig. 5. Moving vehicle - urban environment, STL, Chan1.

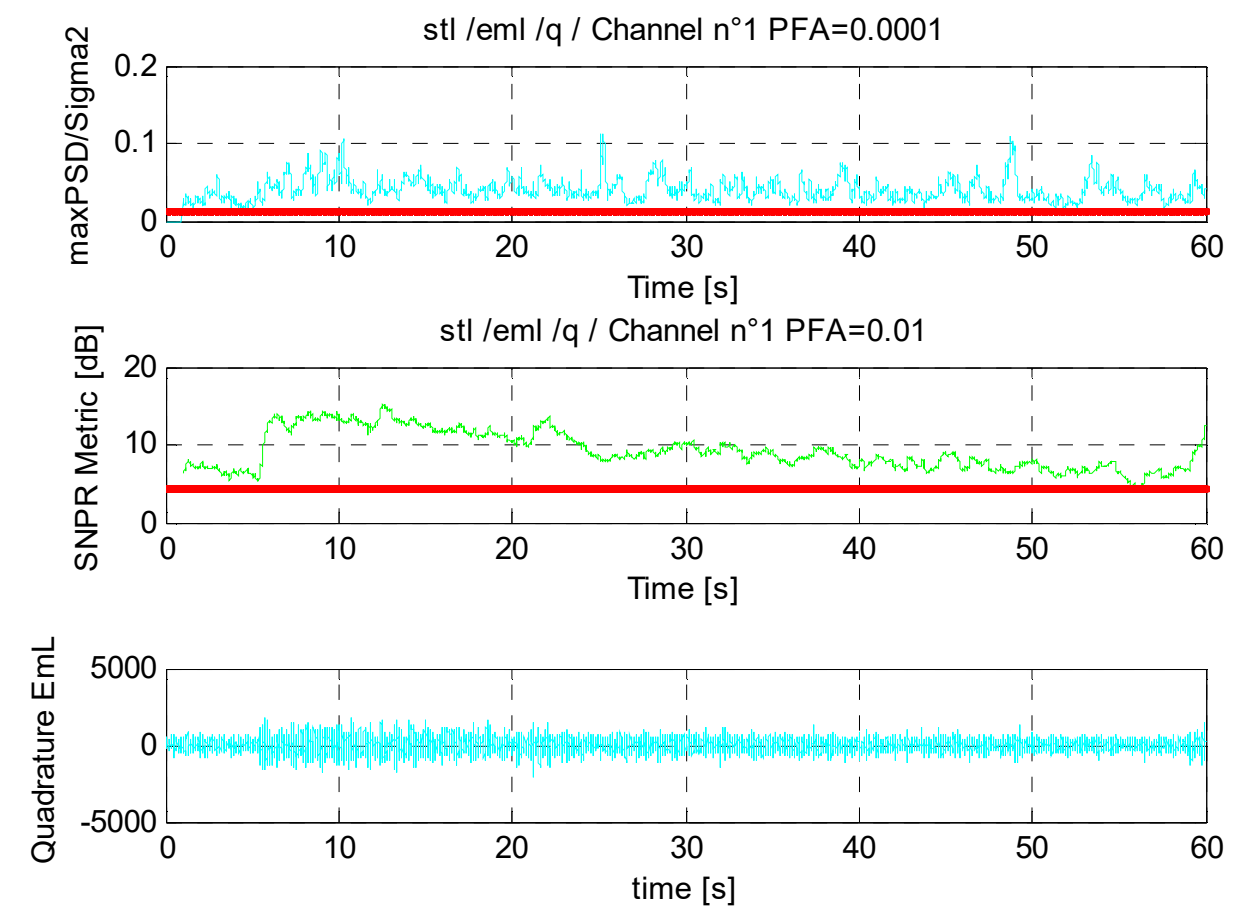

Fig. 6. Moving vehicle - urban environment, STL, Chan1, PFA=$10^{-4}$ (Detector I), PFA=$10^{-2}$ (Detector II).

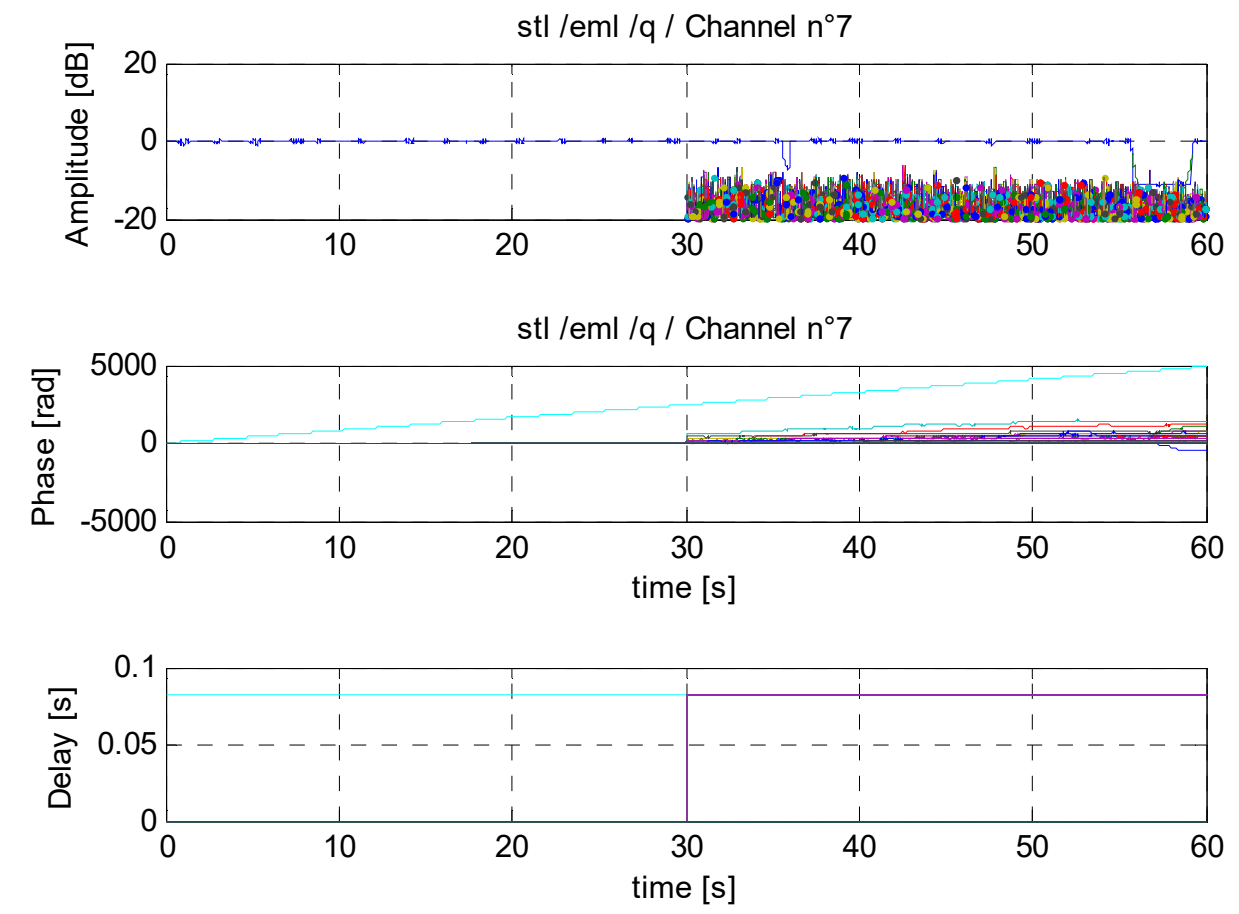

Fig. 7. Moving vehicle - urban environment, STL, Chan7.

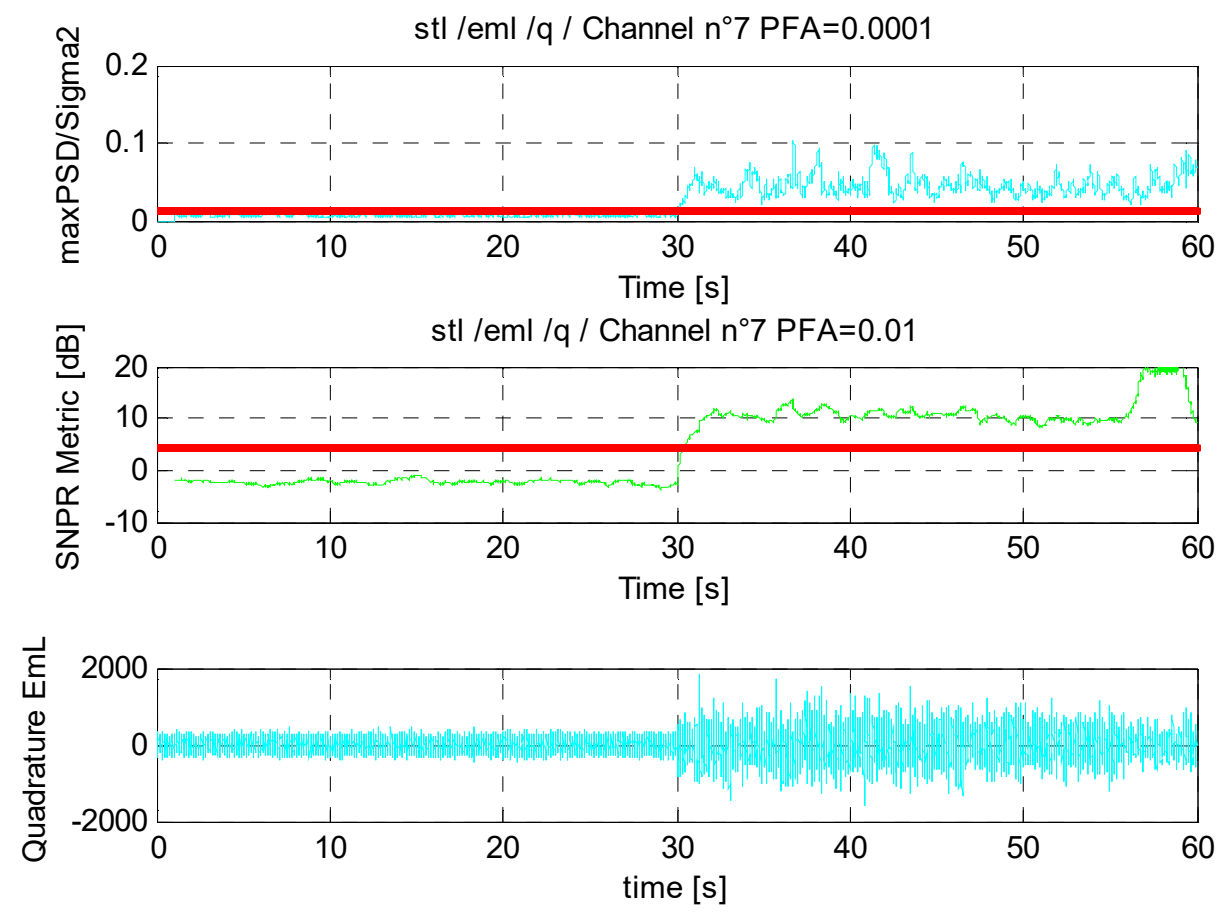

Fig. 8. Moving vehicle - urban environment, STL, Chan7, PFA=$10^{-4}$ (Detector I), PFA=$10^{-2}$ (Detector II).

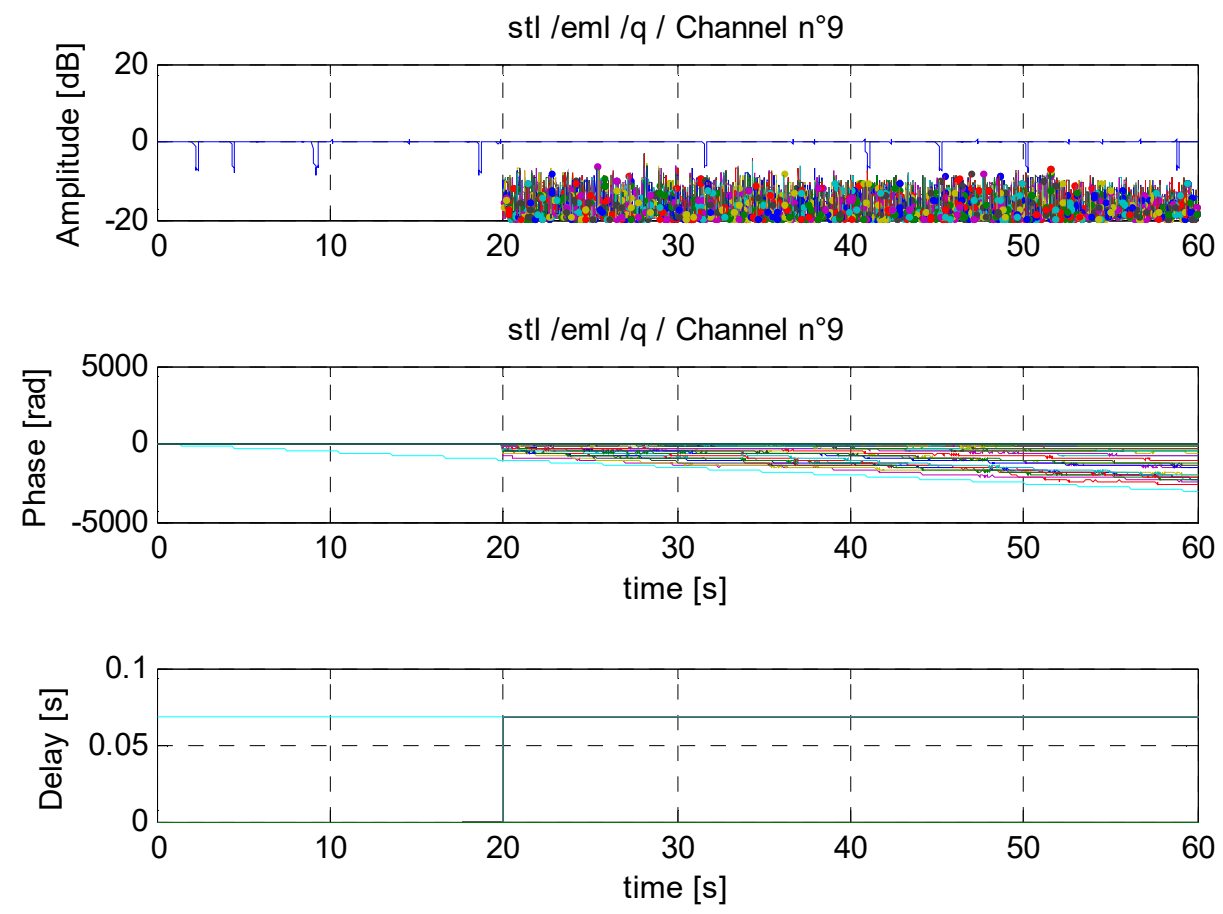

Fig. 9. Moving vehicle - urban environment, STL, Chan9.

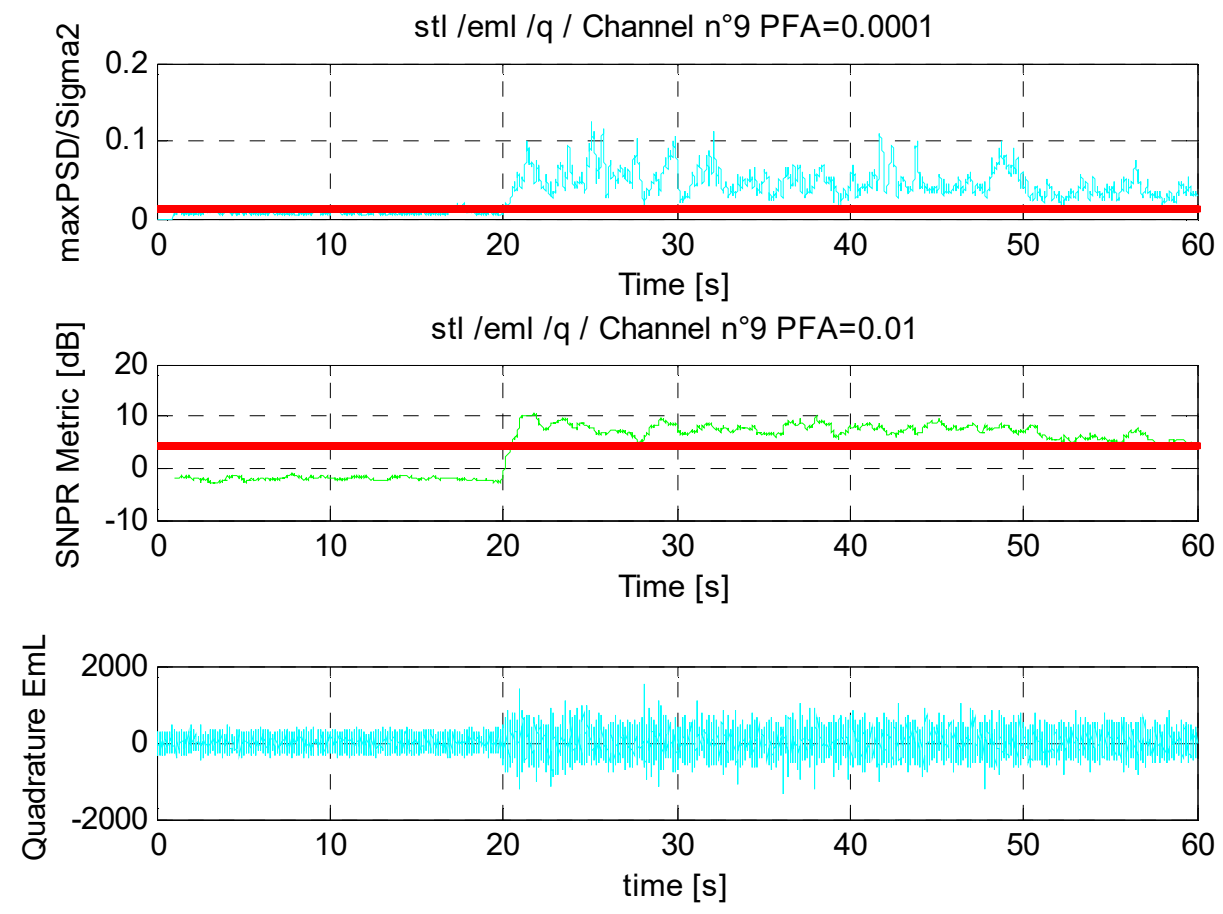

Fig. 10. Moving vehicle - urban environment, STL, Chan9, PFA=$10^{-4}$ (Detector I), PFA=$10^{-2}$ (Detector II).

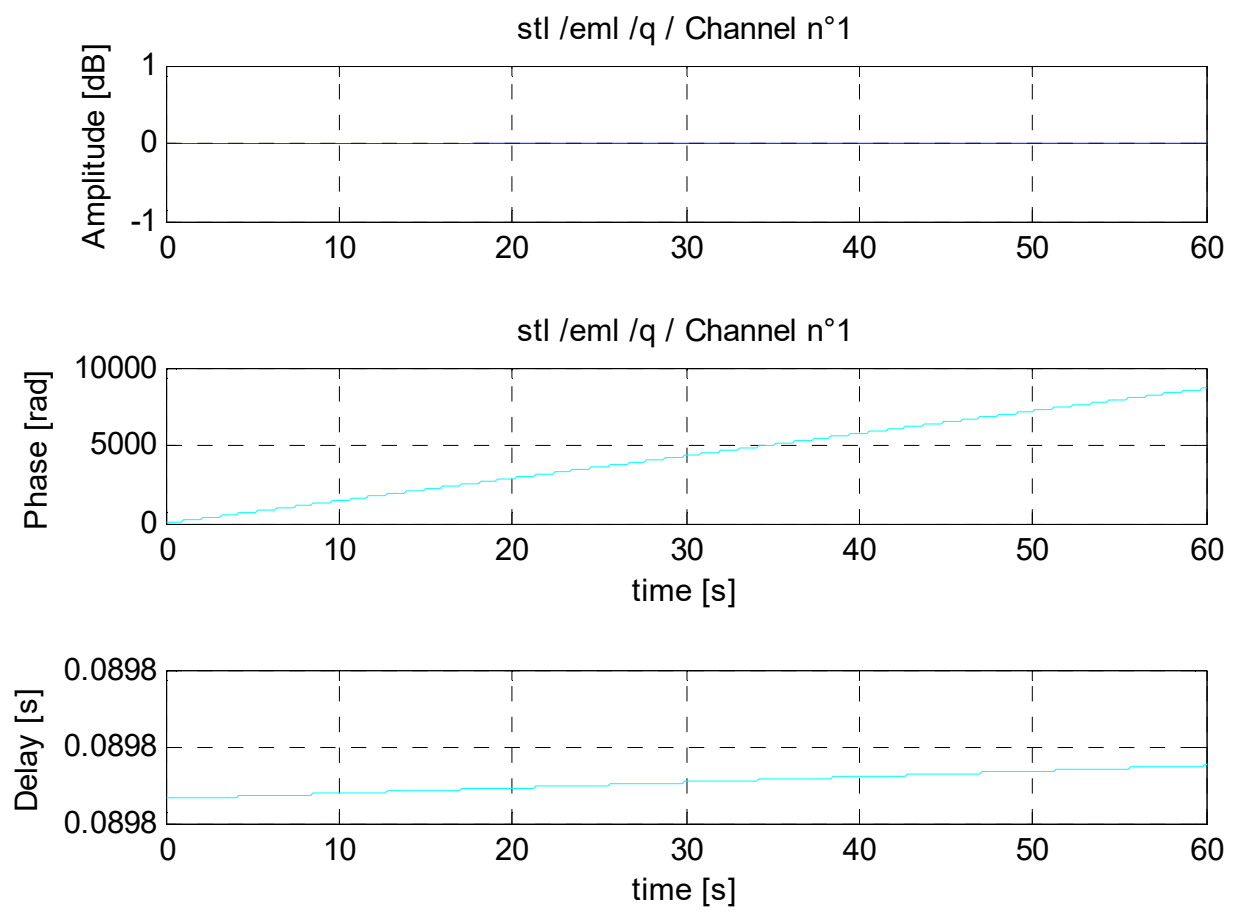

Fig. 11. Moving vehicle – open sky environment, STL, Chan1.

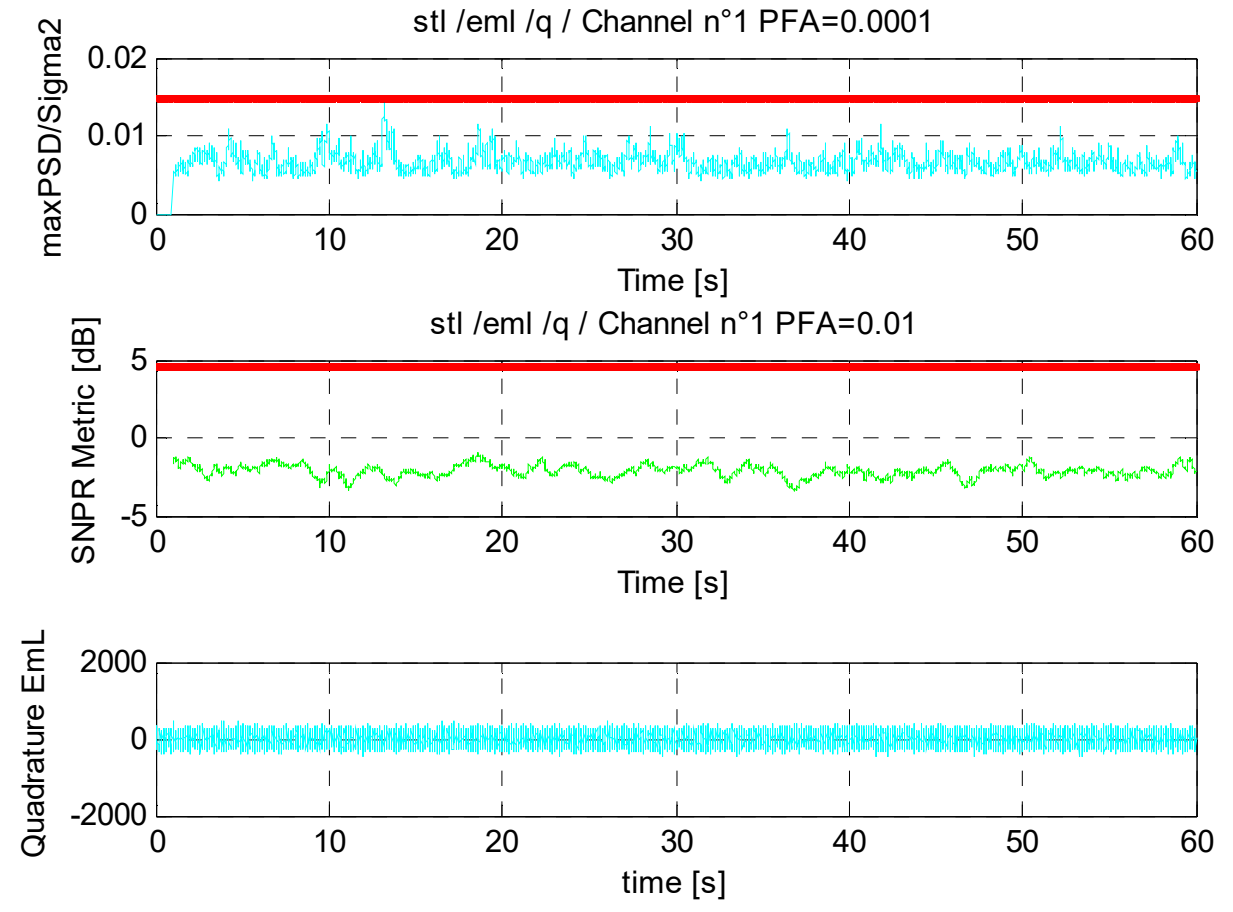

Fig. 12. Moving vehicle – open sky environment, STL, Chan1, PFA=$10^{-4}$ (Detector I), PFA=$10^{-2}$ (Detector II).

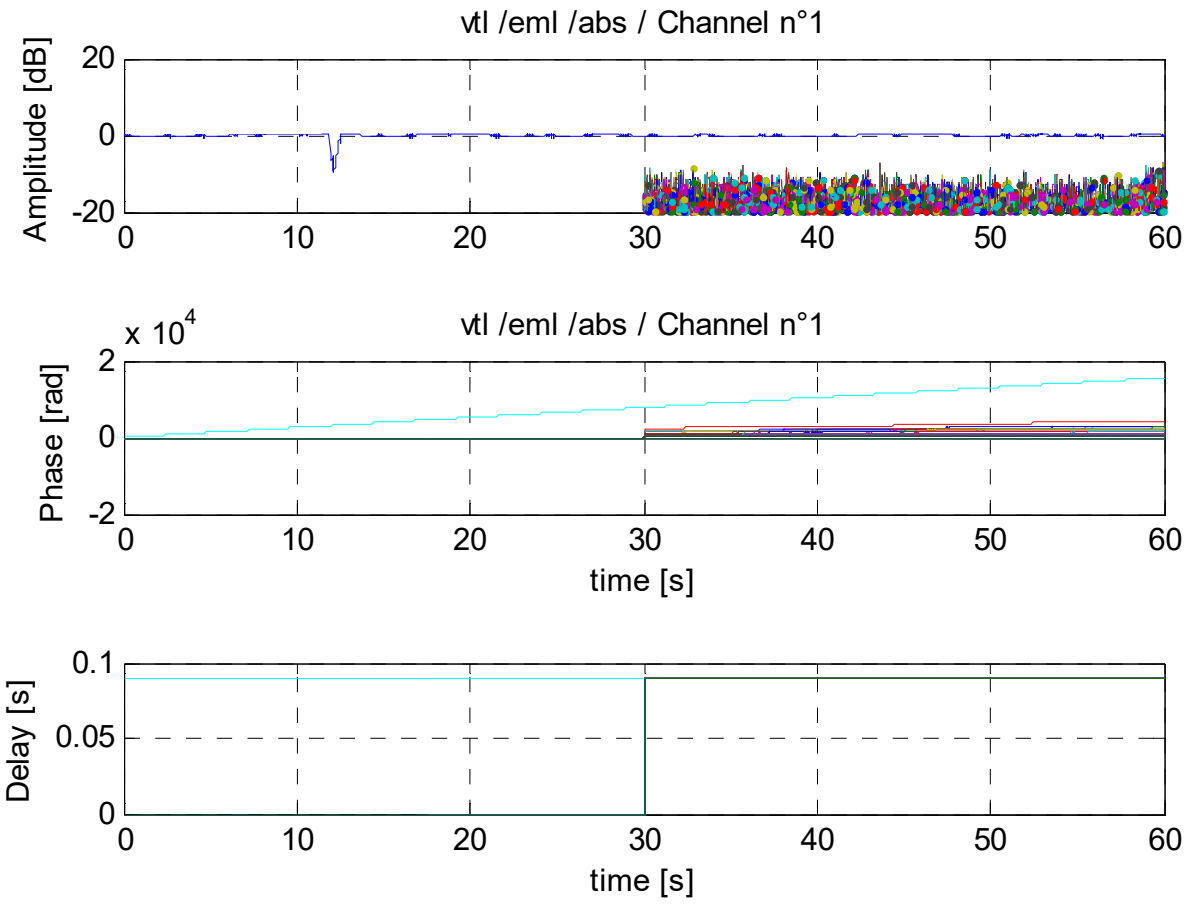

Fig. 13. Moving vehicle - urban environment, VTL, Chan1.

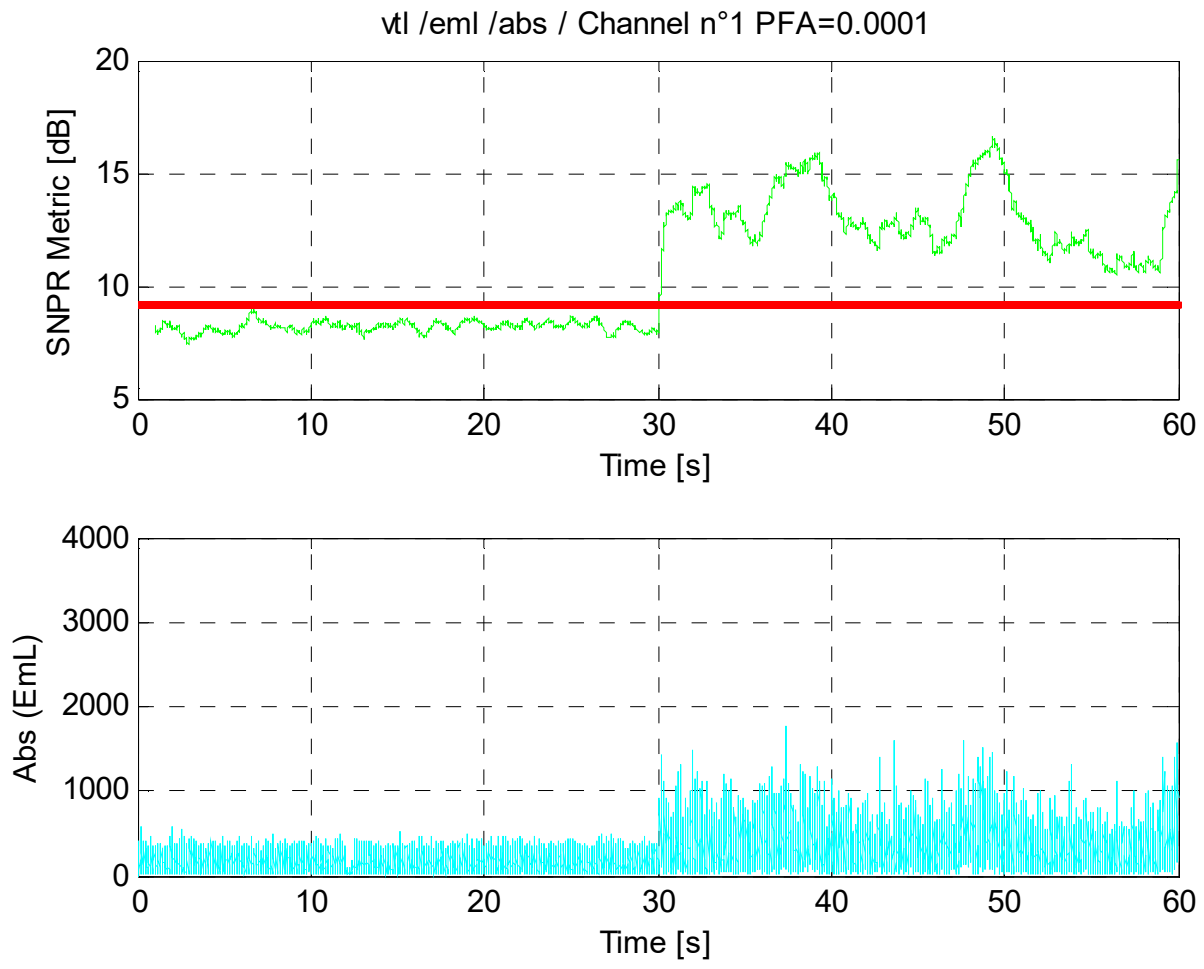

Fig. 14. Moving vehicle - urban environment, VTL, Chan1, PFA=$10^{-4}$ (Detector II).

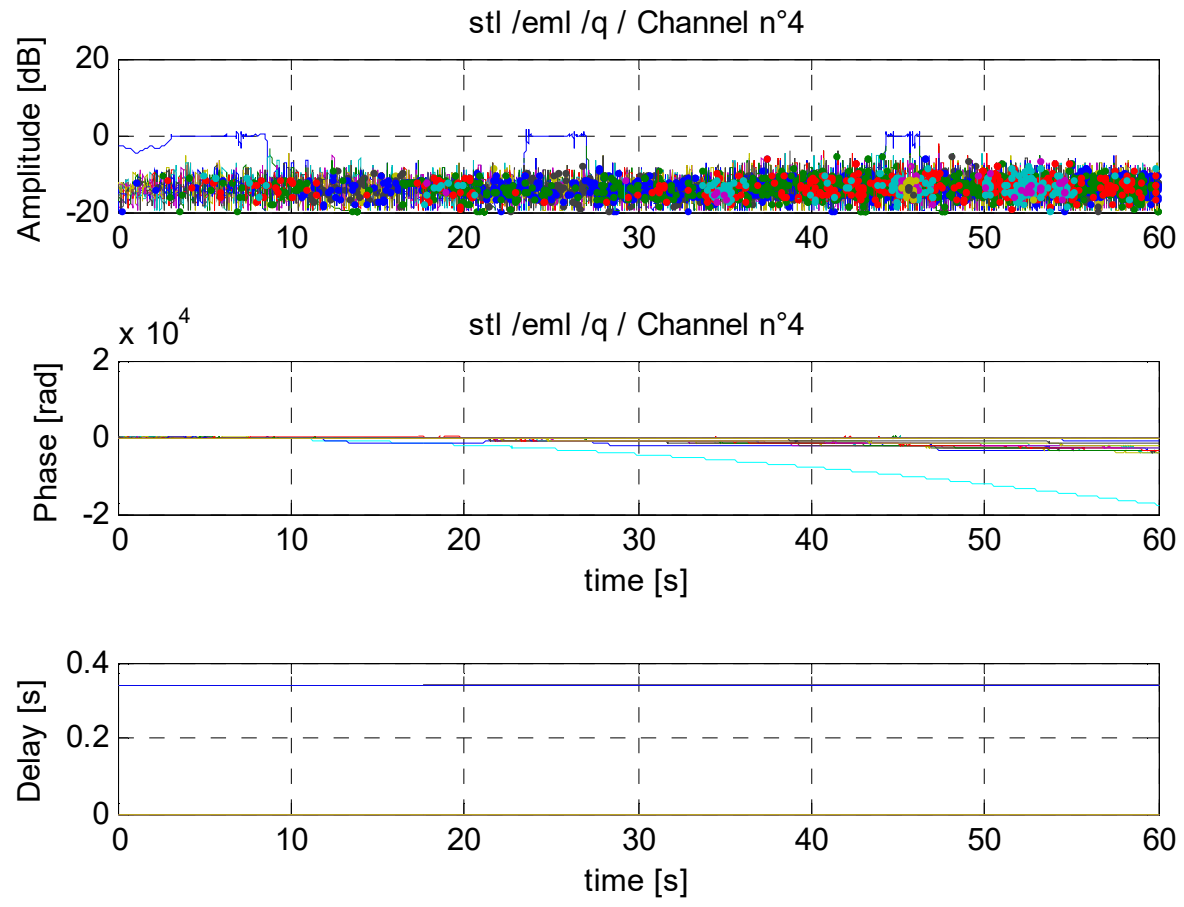

Fig. 15. Moving pedestrian - urban environment, Chan4.

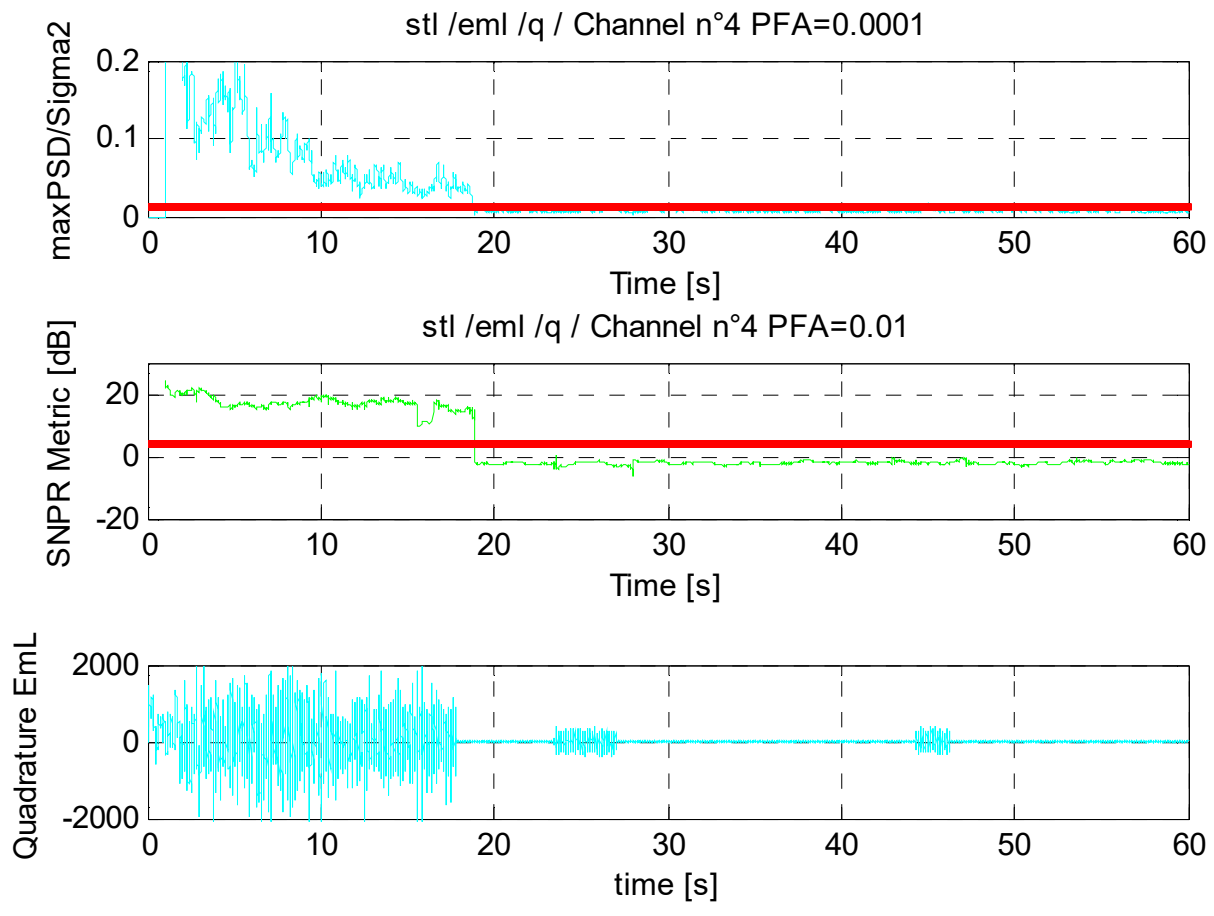

Fig. 16. Moving pedestrian - urban environment, STL, Chan4, PFA=$10^{-4}$ (Detector I), PFA=$10^{-2}$ (Detector II)

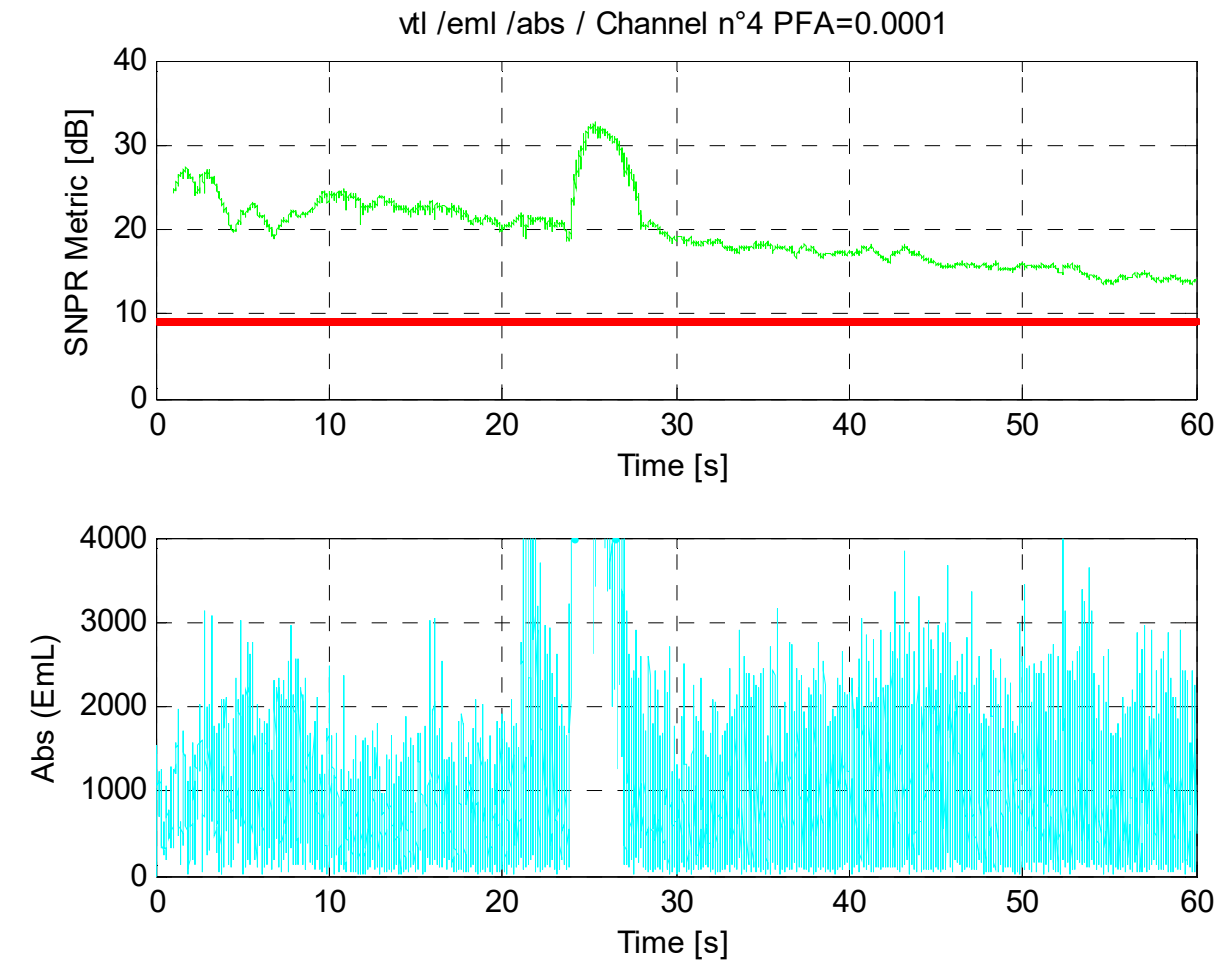

Fig. 17. Moving pedestrian - urban environment, VTL, Chan4, PFA=$10^{-4}$ (Detector II)

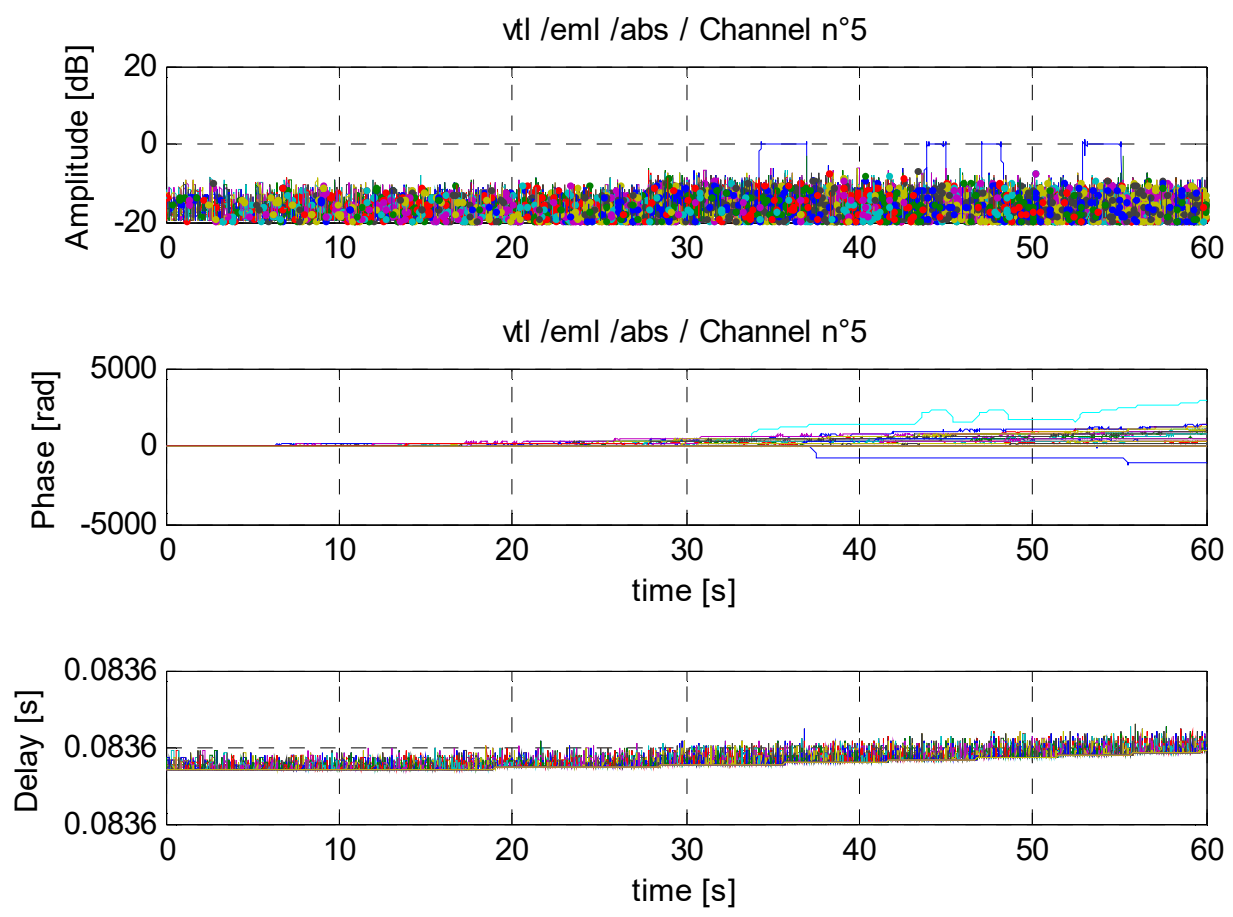

Fig. 18. Moving pedestrian - urban environment, Chan5.

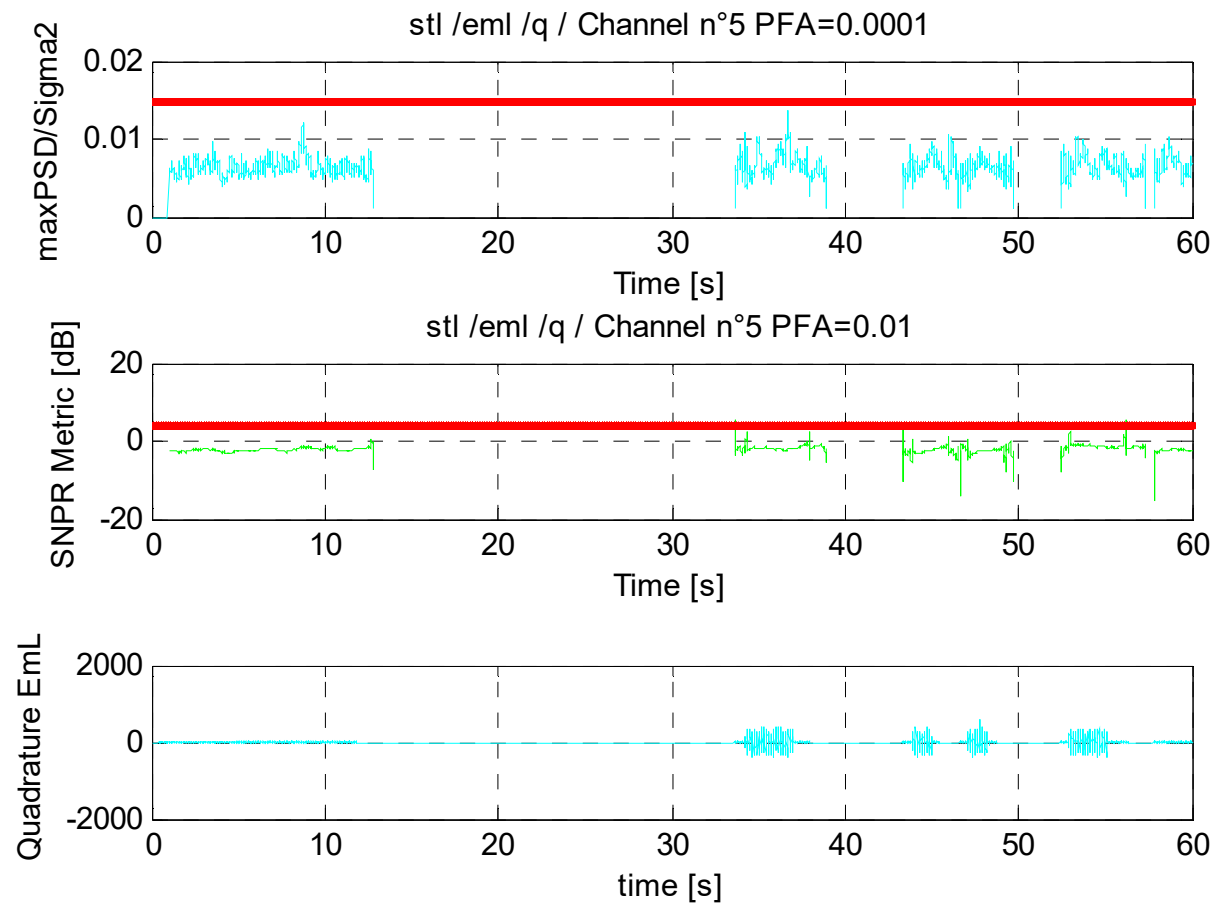

Fig. 19. Moving pedestrian - urban environment, STL, Chan5, PFA=$10^{-4}$ (Detector I), PFA=$10^{-2}$ (Detector II)

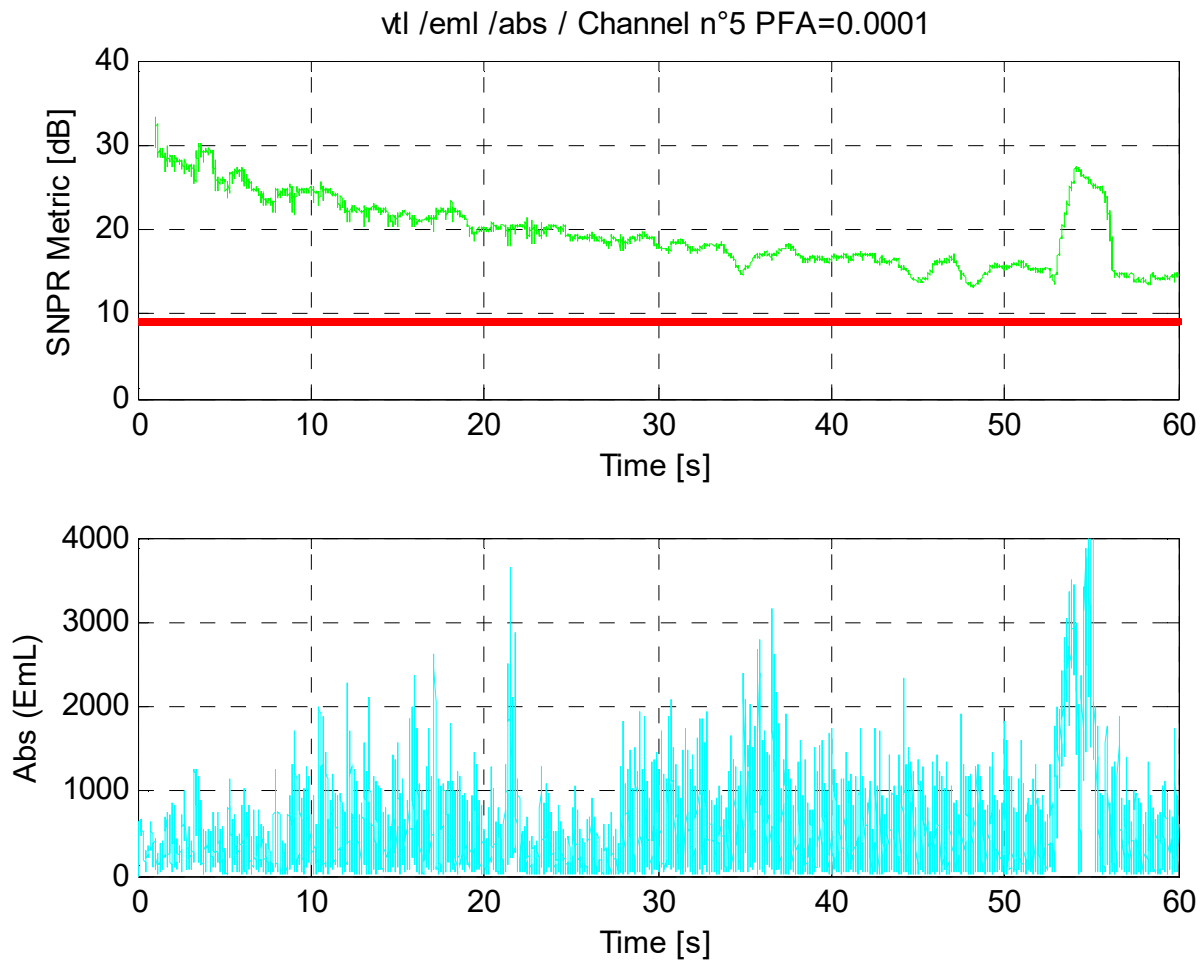

Fig. 20. Moving pedestrian - urban environment, VTL, Chan5, PFA=$10^{-4}$ (Detector II)

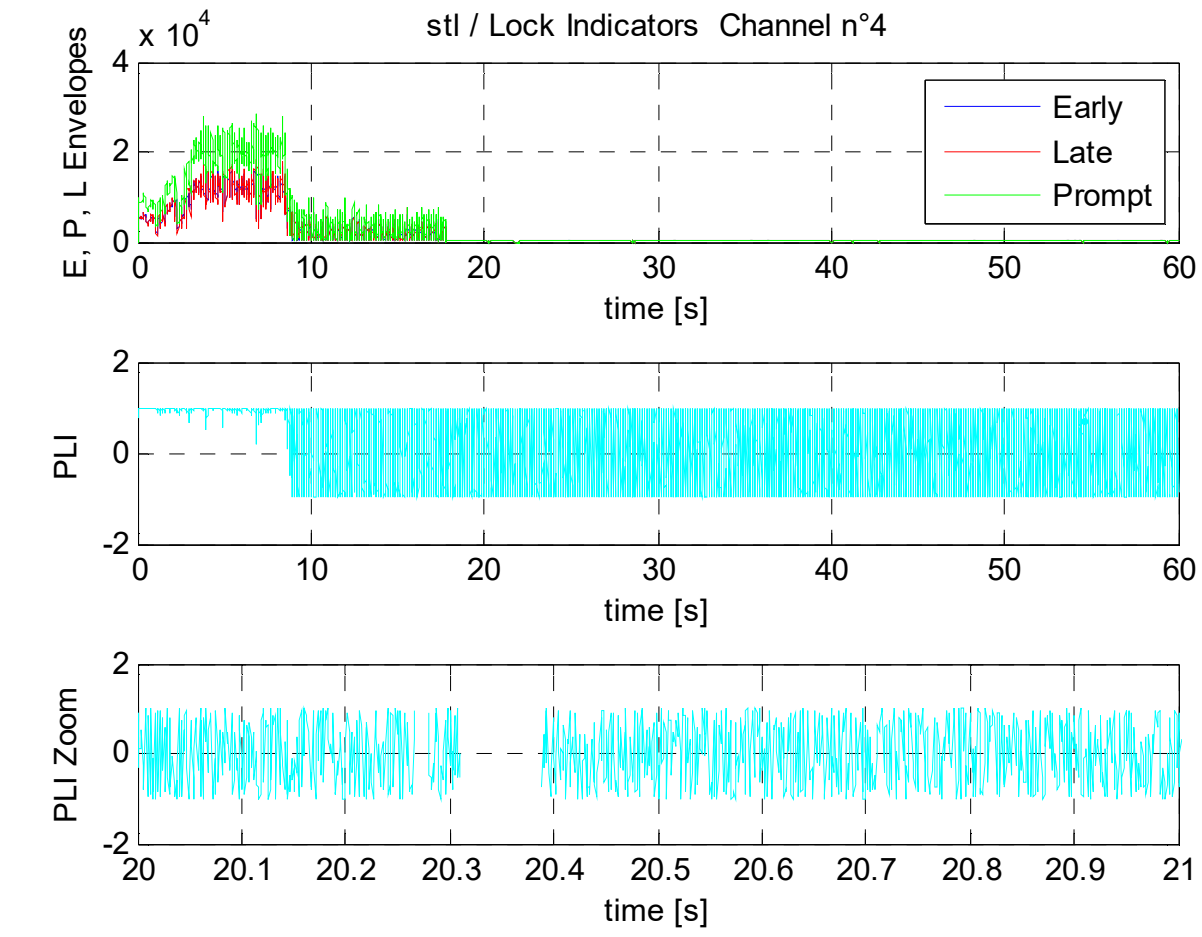

Fig. 21. Moving pedestrian - urban environment, STL, Chan4, Lock Indicators

### *B. Real GPS Signals Data*

In this section, the performance of one of the proposed multipath detection techniques, Detector I, is also evaluated on real GPS signals collected in urban, semi-urban, and foliage environments, for a STL. Digitized signals from the radio frequency (RF) front-end of different GPS receivers are used. A receiver with a high gain antenna provides signals with high carrier to noise ratio ($C/N_0$). A high $C/N_0$ results in a high post-correlation SNR which increases the probability of multipath detection (PD) and theoretically the performance of Detector I as demonstrated in section IV, Fig. 3. The receivers are calibrated to generate digitized signals at different Intermediate Frequencies (IF). Signals that are digitized with different sampling frequencies ($f_s$) are used. The acquisition, code and carrier tracking algorithms, as well as the proposed detection algorithm are implemented using these signals in Matlab. The coherent integration time used during tracking for all signals is 1ms. This means that if a sliding window of N=64 samples is used to set the detection test, 64ms time delay results before detection just after system initialisation.

With the signals synthesized by the DLR multipath simulator, it is possible to control at what instant during tracking the multipath signals can be superimposed on the LOS signal. But with real signals, this is not possible as the provided digitized signal is a mixture of the LOS and multipath signals already. The signals were just collected in environments that are prone to generate multipath contaminated signals.

Figures 22 to 24 are a case of a moving vehicle in a heavy foliage environment, where the detection test threshold is set with PFA=$10^{-1}$, PFA=$10^{-3}$ and PFA=$10^{-5}$. The tracking mode is scalar (STL). A tracking duration of 20 seconds is studied. This is a case of a weak signal (low $C/N_0$). The probability of multipath detection with a low PFA is low for a low post-correlation SNR. Figure 24 represents the same channel as Fig. 22 but compares the signal power on the quadrature arm of EmL with that of the in-phase prompt arm. The figures show that PFAs of $10^{-3}$ or $10^{-5}$ (which are appropriate PFA values to avoid false detections) fail to detect multipath on channels 1 and 2 which are tracking the GPS satellites SV1 and SV20 respectively. With a PFA of 0.1 however, multipath is detected on both channels. A PFA of 0.1 is not recommended because this results in several false detections. It appears however that for a very weak signal, only a high PFA achieves multipath detection for a STL. Depending on whether the number of visible satellites at a given time is enough to compute a navigation solution, a high PFA can be chosen in order to increase the chance of excluding all multipath affected satellites, and a lower PFA may be chosen to more likely exclude only the ones that are severely affected. Figures 25 to 27 show scenarios of a moving vehicle in a semi-urban environment, and the test is set with PFA=$10^{-1}$, PFA=$10^{-3}$ and PFA=$10^{-5}$. The $C/N_0$ is high enough to get multipath detections with PFA=$10^{-3}$ even with PFA=$10^{-5}$ in some cases on channels 4 and 7 which are tracking satellites SV22 and SV3 respectively.

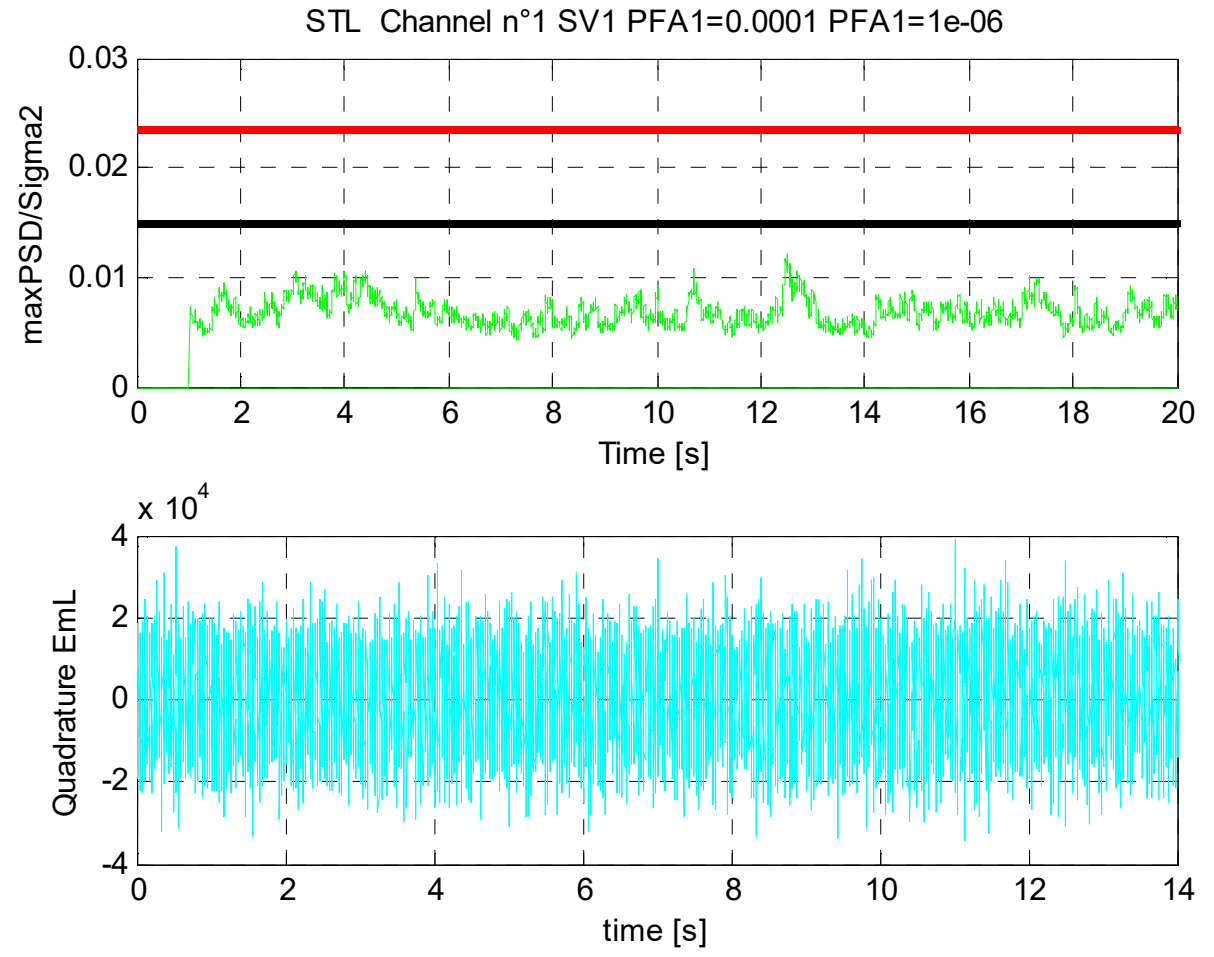

Fig. 22. Moving vehicle – foliage environment, STL, SV1, low $C/N_0$, IF = 0 MHz, $f_s$ = 26 MHz, Detector I.

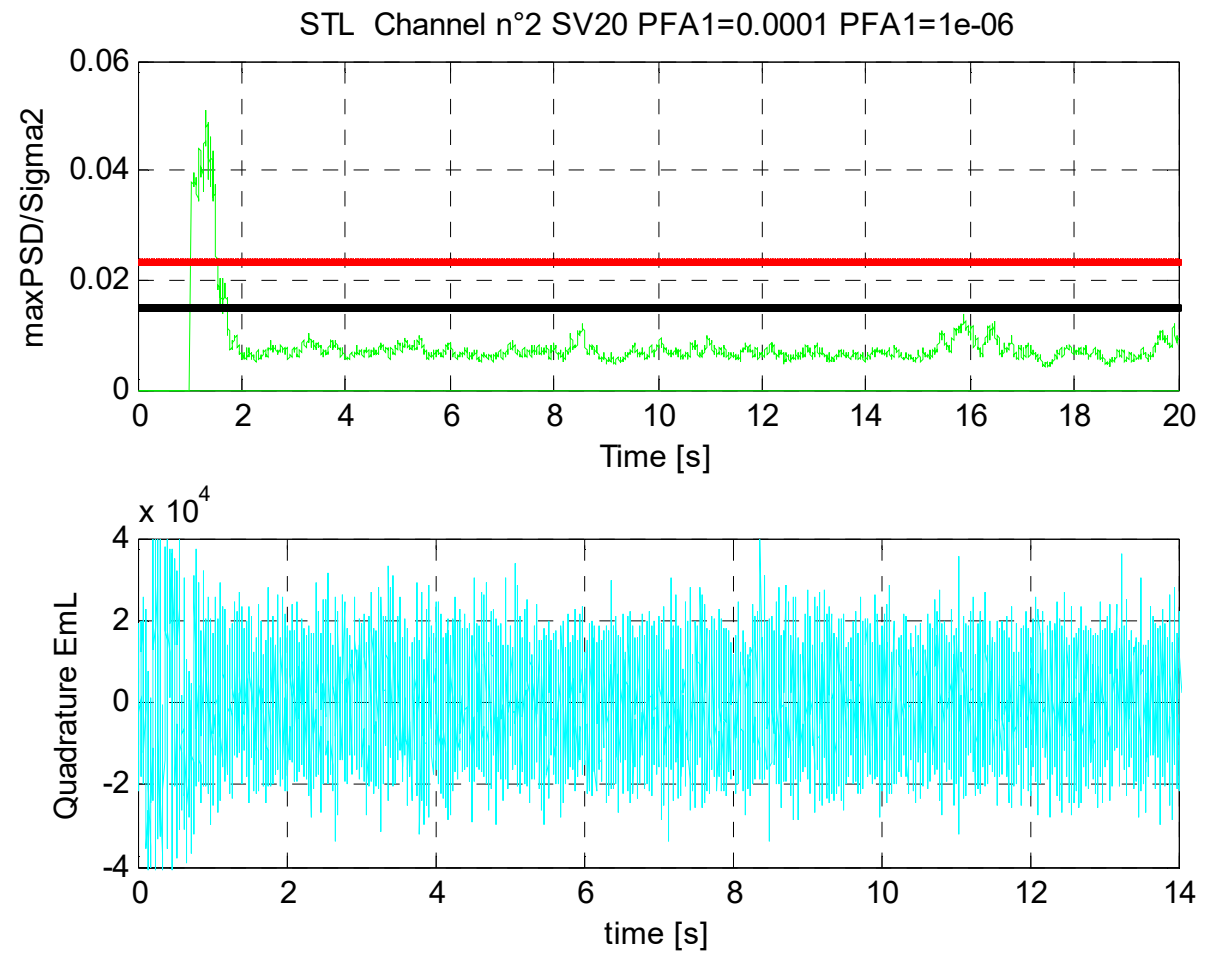

Fig. 23. Moving vehicle – foliage environment, STL, SV20, low $C/N_0$, IF = 0 MHz, $f_s$ = 26 MHz, Detector I.

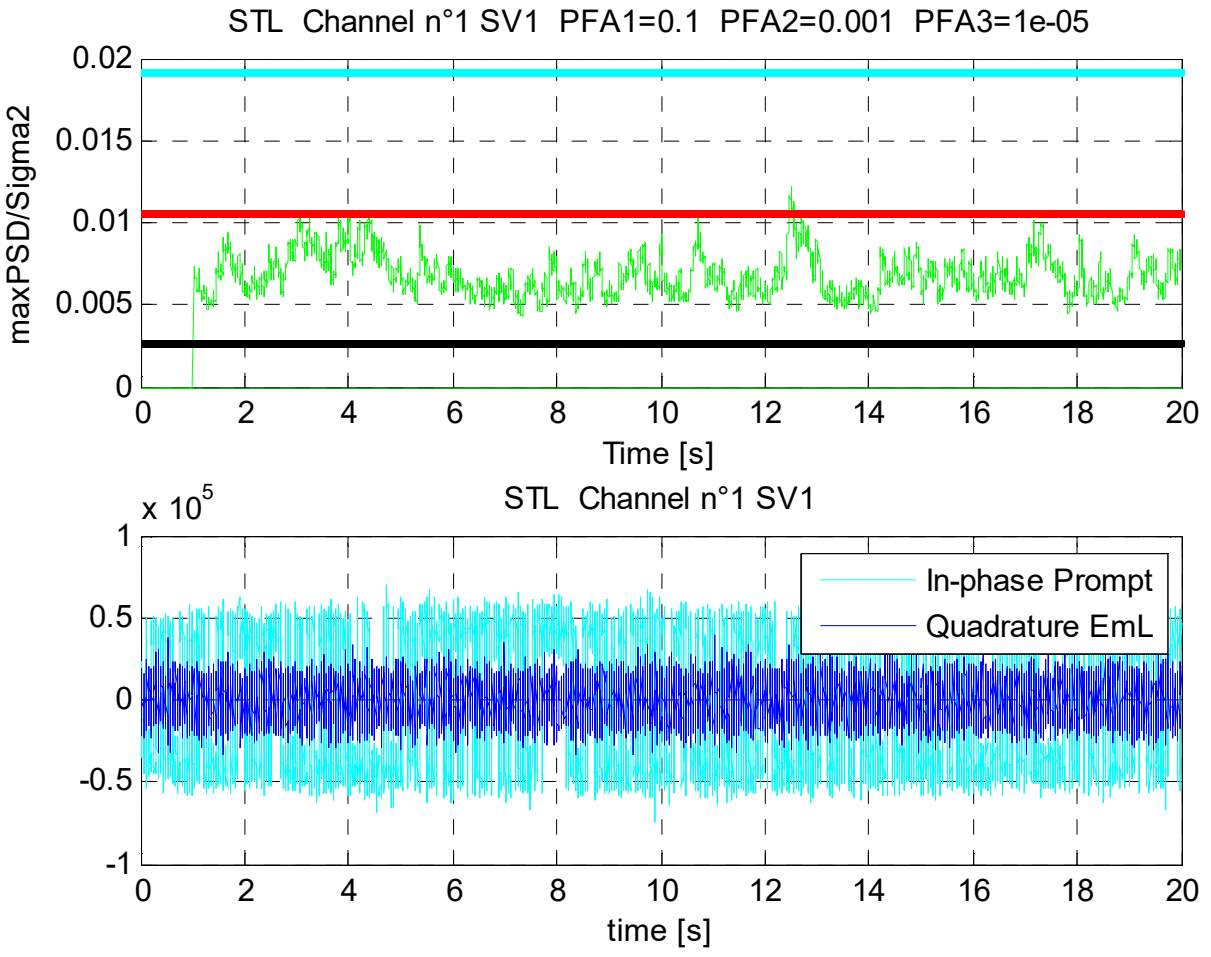

Fig. 24. Moving vehicle – foliage environment, STL, SV1, low $C/N_0$, IF = 0 MHz, $f_s$ = 26 MHz, Detector I.

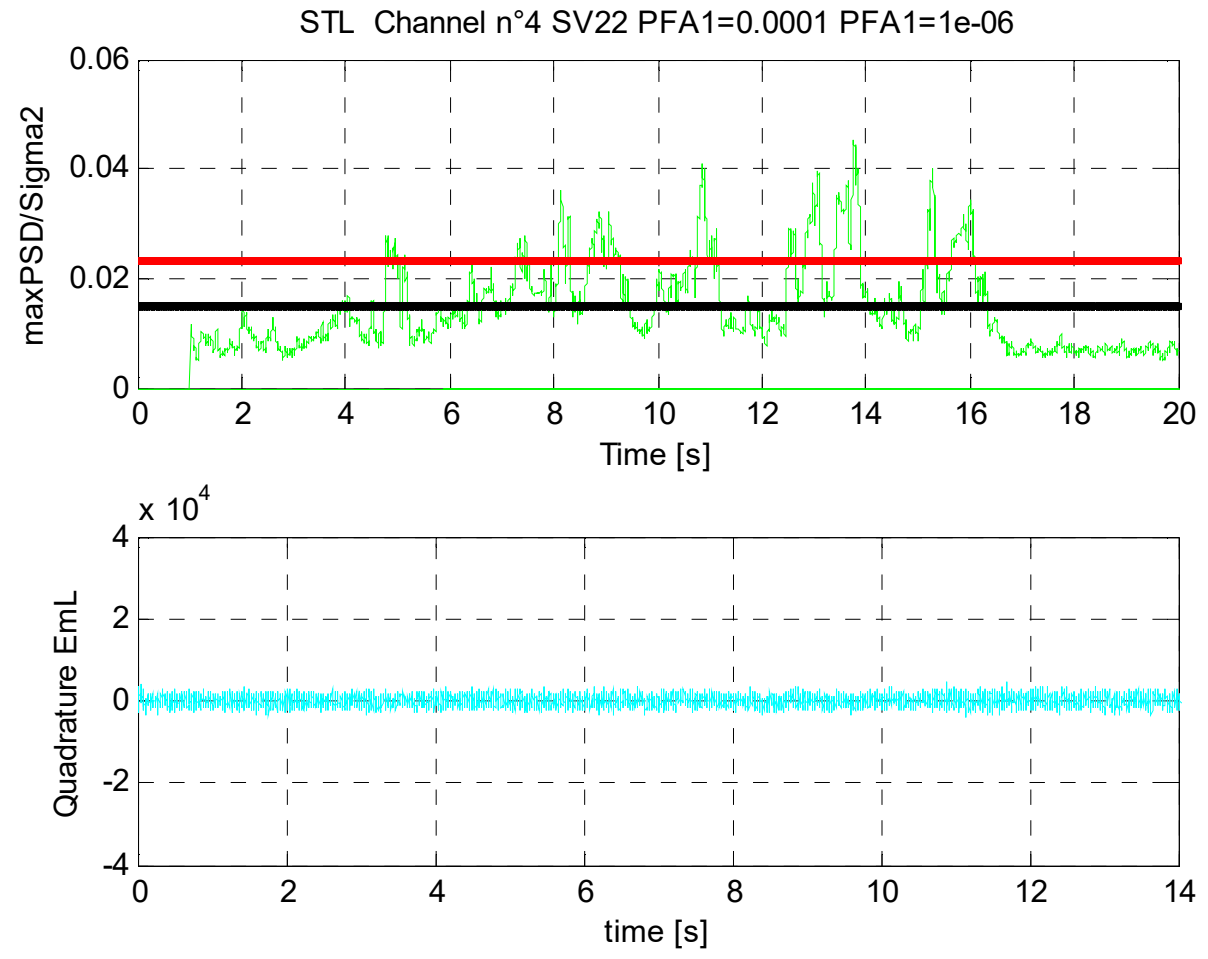

Fig. 25. Moving vehicle – semi-urban environment, STL, SV22, high $C/N_0$, IF = 9.548 MHz, $f_s$ = 38.192 MHz, Detector I.

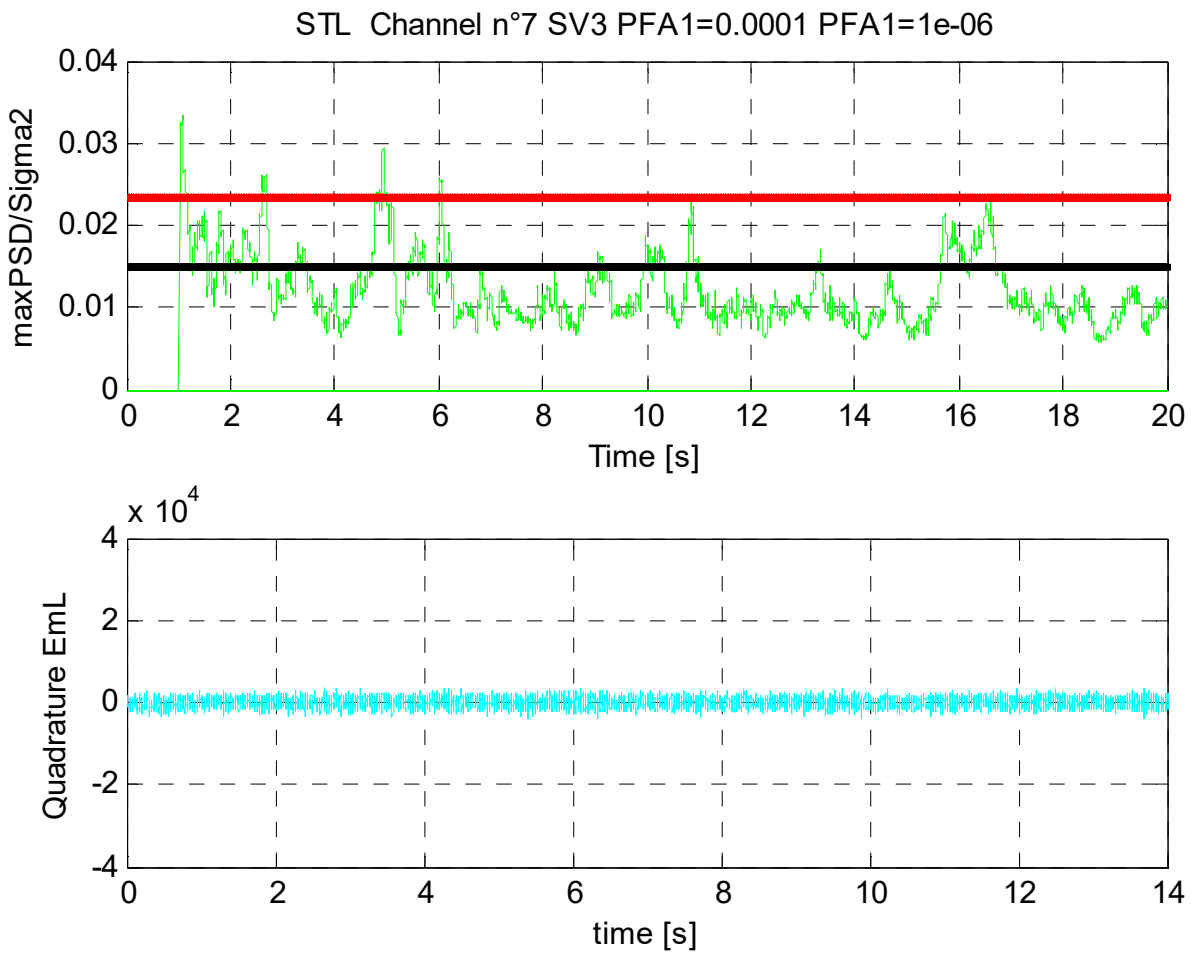

Fig. 26. Moving vehicle – semi-urban environment, STL, SV3, high C/$N_0$, IF = 9.548 MHz, $f_s$ = 38.192 MHz, Detector I.

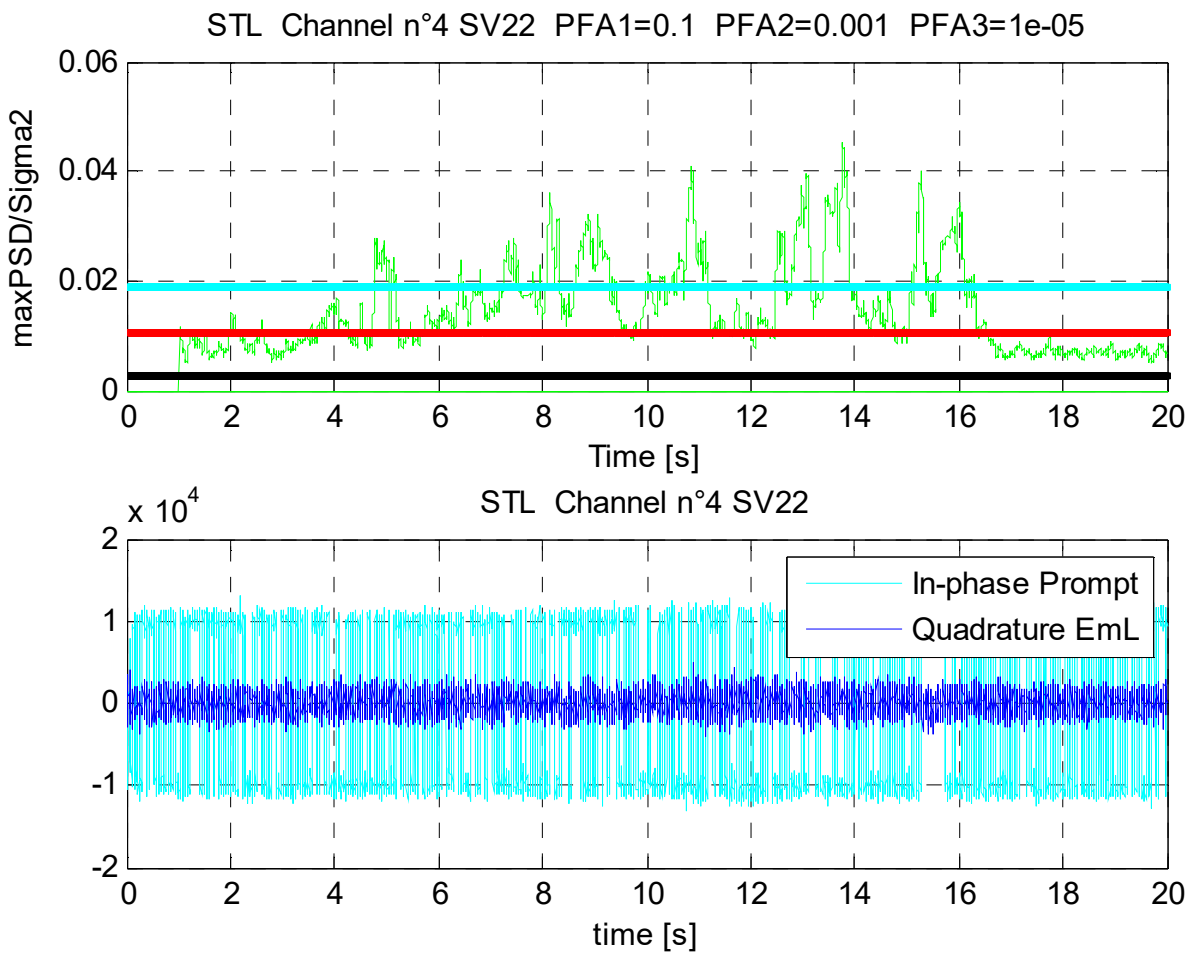

Fig. 27. Moving vehicle – semi-urban environment, STL, SV22, high C/$N_0$, IF = 9.548 MHz, $f_s$ = 38.192 MHz, Detector I.

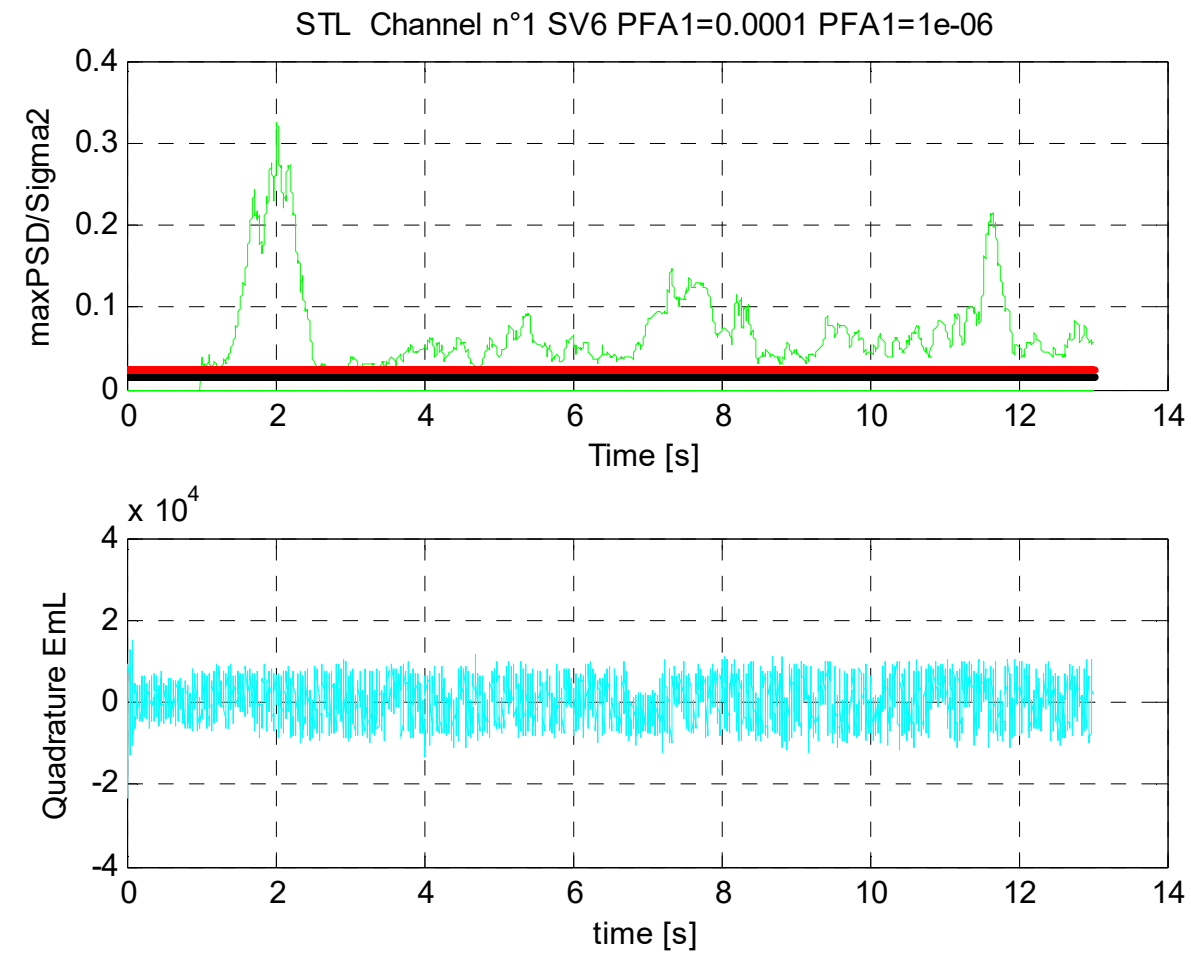

Fig. 28. Fixed point – urban environment, STL, SV6, high C/$N_0$, IF = 1.25 MHz, $f_s$ = 5 MHz, Detector I.

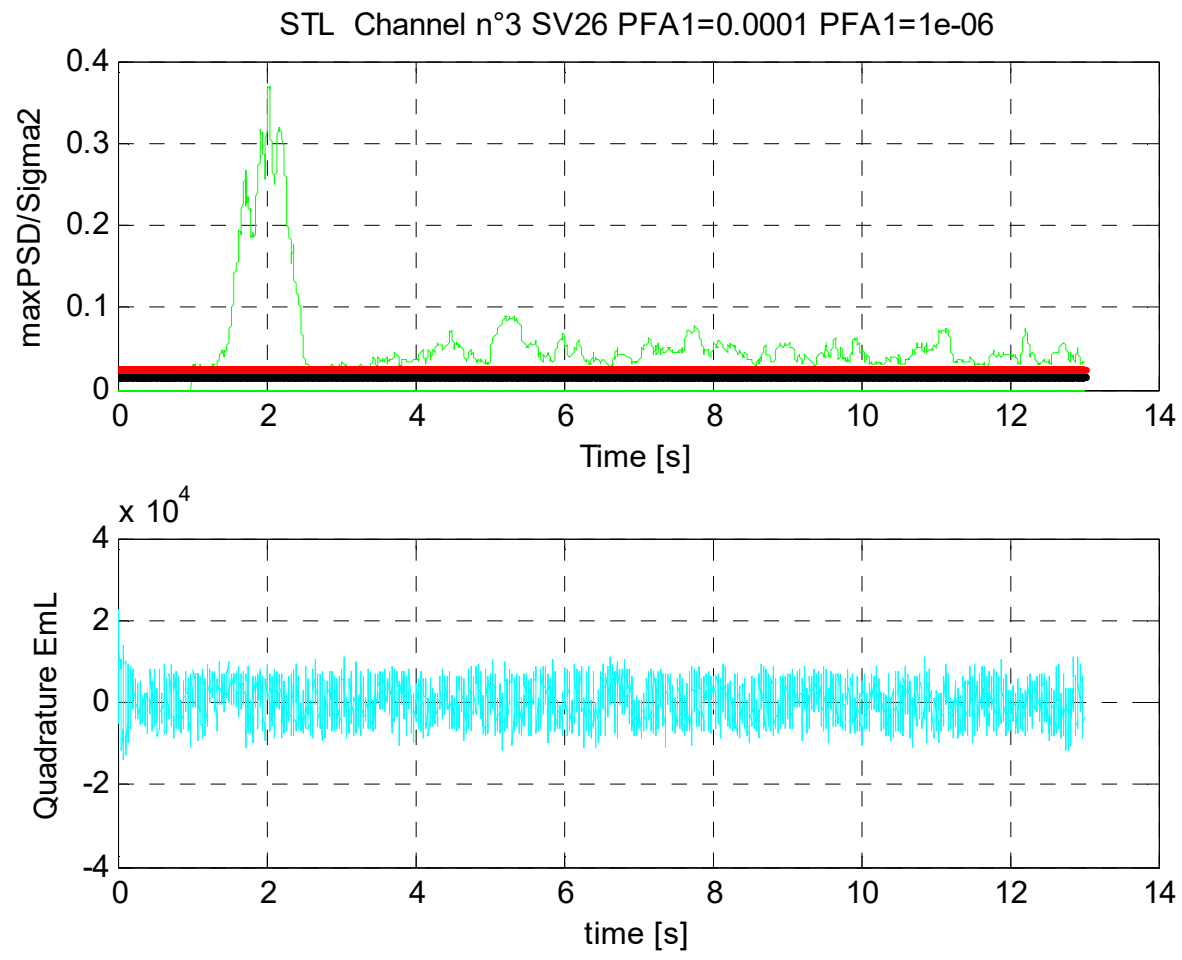

Fig. 29. Fixed point – urban environment, STL, SV26, high C/$N_0$, IF = 1.25 MHz, $f_s$ = 5 MHz, Detector I.

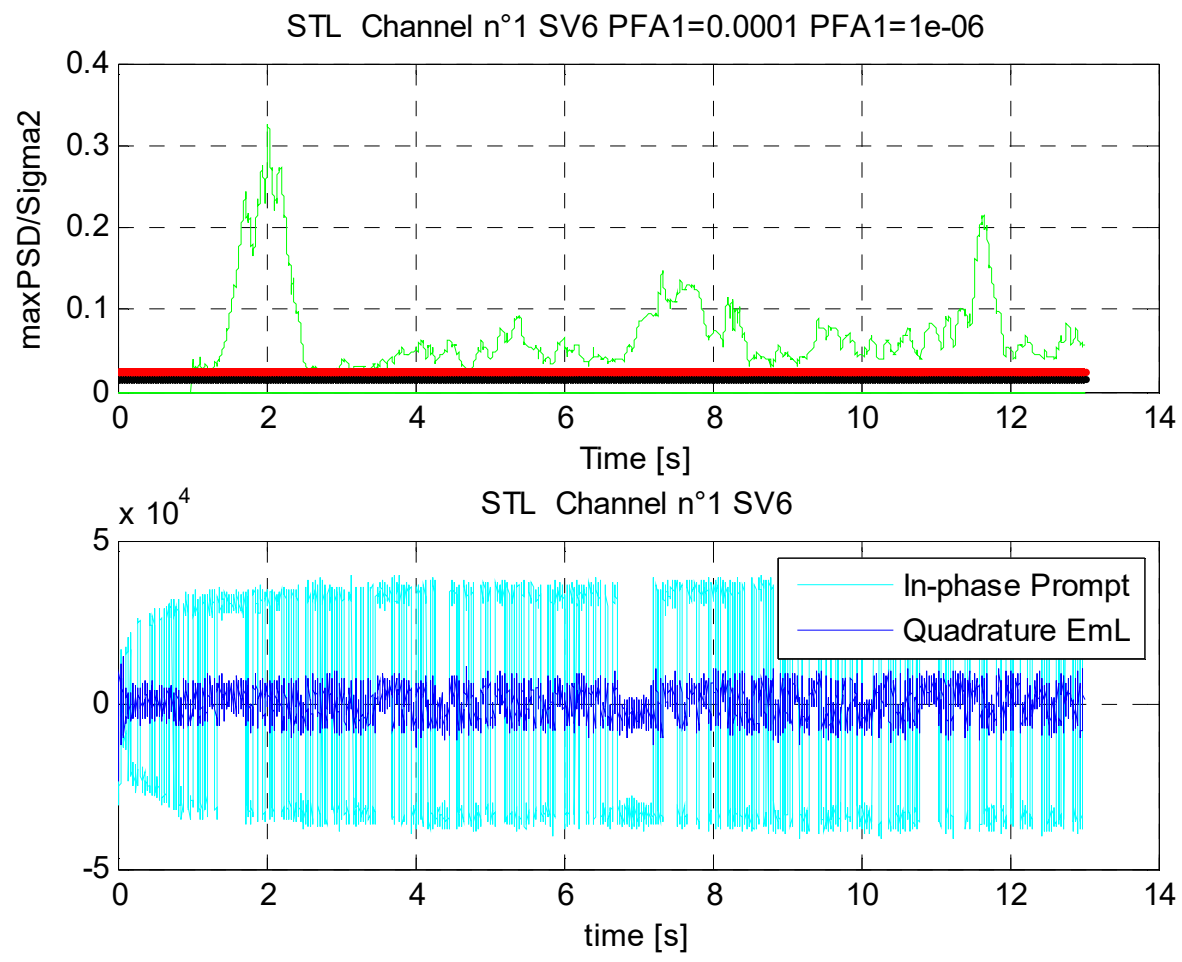

Fig. 30. Fixed point – urban environment, STL, SV6, high C/$N_0$, IF = 1.25 MHz, $f_s$ = 5 MHz, Detector I.

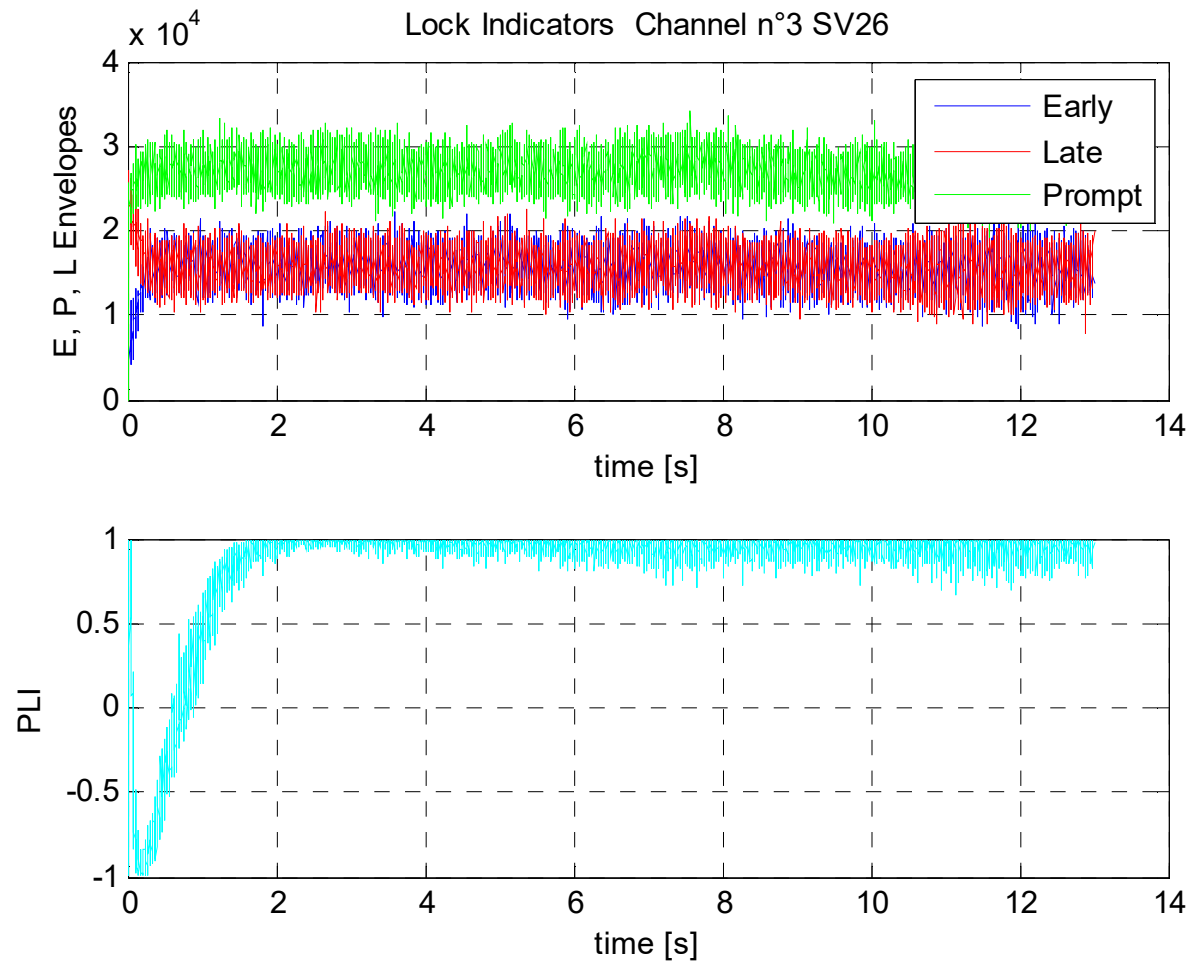

Fig. 31. Fixed point – urban environment, STL, SV26, high C/$N_0$, IF = 1.25 MHz, $f_s$ = 5 MHz, Lock Indicators

Figure 28 to 30 show the results of a fixed point in an urban environment. The processed signal has high $C/N_0$. PFAs of $10^{-4}$ and $10^{-6}$ which are low enough to avoid false detections are used. Multipath is detected almost for the whole 13 seconds tracking duration on channels 1 and 3 tracking satellites SV6 and SV26 respectively. Figure 31 illustrates the delay and phase lock indicators for the scenario of a fixed point in urban environment of Fig. 29.

## VI. Conclusion

This paper has suggested two multipath detection techniques that are based on FFT. Application of the detection techniques to synthesized and real GPS signals on correlator outputs of the code and carrier tracking loops (STL or VTL) demonstrate their efficiency in detecting multipath contaminated tracking channels. The detectors are blind to multipath signals that arrive in phase or opposition of phase with the LOS signal for the STL case. It has been demonstrated that when the multipath signal arrives aligned with the LOS signal, i.e. in phase (relative phase of 0° + k360°) or opposition of phase (relative phase of 180° + k360°), the $Q_{EmL}$ output remains blind to multipath presence. This means that for those multipath phase values, the $Q_{EmL}$ signal power does not change according to the presence or absence of multipath and only the post correlation noise is observed on the $Q_{EmL}$ arm. This results in missed detection if the multipath remains in phase or opposition of phase for the duration of the N-sample size window used to set the detection test. This blindness to multipath that is in phase or opposition of phase with the LOS is however not observed in the VTL case. In VTL, when $Q_{EmL}$ is blind to multipath $I_{EmL}$ is not and vice versa. As both $I_{EmL}$ and $Q_{EmL}$ are used in the VTL detection metric, the detectors performance is not affected. These detectors can be utilized as tools to exclude multipath contaminated satellites from the calculation of the PVT solution and to switch between STL and VTL tracking modes depending on accuracy and availability needs for a receiver embedded with both tracking modes. Also, the tests can be applied with minor adjustments to other GNSS systems such as Galileo, GLONASS and Beidou.

## References

[1] S. Kay, Fundamentals of statistical signal processing, Volume II: Detection theory, Prentice Hall, 1998.

[2] O. Mubarak and A. Dempster, "Exclusion of Multipath-Affected Satellites Using Early Late Phase," *Journal of Global Positioning Systems,* vol. 9, no. 2, pp. 145-155, 2010.

[3] O. Mubarak, "Analysis of Early Late Phase for Multipath Mitigation," *ION GNSS 21st. International Technical Meeting of the Satellite Division,* pp. 669-678, 16-19 September 2008.

[4] M. Brenneman, Y. Morton and Q. Zhou, "An ANOVA-Based GPS Multipath Detection Algorithm Using Multi-Channel Software Receivers," in *proceedings of 2008 IEEE/ION Position, Location and Navigation Symposium*, 2008.

[5] Y. Fang, Y. Hong, O. Zhou, W. Liang and L. WenXue, "A GNSS Satellite Selection Method Based on SNR Fluctuation in Multipath Environments," *International Journal of Control and Automation,* vol. 8, no. 11, pp. 313-324, 2015.

[6] A. Beitler, A. Tollkuehn, D. Giustiniano and B. Plattner, "CMCD: Multipath Detection for Mobile GNSS Receivers," in *proceedings of 2015 International Technical Meeting of The Institute of Navigation*, Dana Point, California, January 26 - 28, 2015.

[7] A. Dierendonck, P. Fenton and T. Ford, "Theory and Performance of Narrow Correlator Spacing in a GPS Receiver," *NAVIGATION, Journal of the Institute of Navigation,* vol. 39, no. 3, pp. 265-283, 1992.

[8] V. Veitsel, A. Zhdanov and M. Zhodzicshky, "The Mitigation of Multipath Errors by Strobe Correlators in GPS/GLONASS Receiver," *GPS Solutions,* vol. 2, no. 2, pp. 38-45, 1998.

[9] G. McGraw, "Practical GPS Carrier Phase Multipath Mitigation using High Resolution Correlator Techniques," *Proc. IAIN World Congress/ION Annual Meeting,* pp. 373-381, 2000.

[10] G. McGraw and M. Braasch, "GNSS Multipath Mitigation using Gated and High Resolution Correlator Concepts," *Proc. ION National Technical Meeting,* pp. 333-342, 1999.

[11] B. Townsend and P. Fenton, "A Practical Approach to the Reduction of Pseudorange Multipath Errors in a L1 GPS Receiver," *Proc. 7th International Technical Meeting of the Satellite Division of the Institute of Navigation, Part 1 (of 2), Proc. ION GPS,* vol. 1, pp. 143-148, 1994.

[12] J. Sleewaegen and F. Boon, "Mitigating short-delay multipath: a promising new technique," *Proc. ION GPS,* pp. 204-213, 11 - 14 September 2001.

[13] S. Kay, Fundamentals of statistical signal processing, Volume I: Estimation theory, Prentice Hall, 1993.

[14] R. van Nee, "The Multipath Estimating Delay Lock Loop," *IEEE 2nd International Symposium on Spread Spectrum Techniques and Applications,* pp. 39-42, 1992.

[15] C. Cahn and M. Chansarkar, "Multipath Corrections for a GPS Receiver," *Proc. 10th International Technical Meeting of the Satellite Division of the Institute of Navigation (ION-GPS),* vol. 1, pp. 551-557, 1997.

[16] L. Weil, "Multipath Mitigation using Modernized GPS Signals: How Good Can it Get?," *Proc. 17th International Technical Meeting of the Satellite Division of the Institute of Navigation (ION GPS),* pp. 493-505, 2002.

[17] P. Fenton and J. Jones, "The Theory and Performance of NovAtel Inc.'s Vision Correlator," *Proc. 19th International Technical Meeting of the Satellite Division of the Institute of Navigation (ION GNSS' 05),* p. 2178–2186, September 2005.

[18] M. Sahmoudi and R. Landry, "Multipath Mitigation Techniques Using Maximum-Likelihood Principle," *Inside GNSS,* pp. 24-29, November/December 2008.

[19] D. Kumar and K. Lau, "Deconvolution Approach to Carrier and Code Multipath Error Elimination in High Precision GPS," *Proc. 1996 National Technical Meeting of the Institute of Navigation,* pp. 729-737, January 1996.

[20] S. Lohan, D. Skournetou and A. Sayed, "A Deconvolution Algorithm for Estimating Jointly the Line-Of-Sight Code Delay and Carrier Phase of GNSS Signals," *ENC-GNSS,* 2009.

[21] C. Yang and L. Porter, "Frequency-Domain Characterization of GPS Multipath for Estimation and Mitigation," *Proc. 18th International Technical Meeting of the Satellite Division of the Institute of Navigation (ION-GNSS),* September 2005.

[22] C. Yang and L. Porter, "Multipath-Desensitized Delay Estimation with GPS Signal Channel Transfer Function Filtering," *ION 61st Annual Meeting,* June 2005.

[23] J. Ray, "Mitigation of GPS Code and Carrier Phase Multipath Effects Using a Multi-Antenna System," *PhD Thesis,* March 2000.

[24] V. Heiries, "Optimisation d'une chaîne de réception pour signaux de radionavigation par satellite à porteuse à double décalage (BOC)," *PhD Thesis,* 2007.

[25] Y. Chan, Q. Yuan, H. So and R. Inkol, "Detection of stochastic signals in the frequency domain," *IEEE Transactions on Aerospace and Electronic Systems,* vol. 37, no. 3, pp. 978 - 988, Jul 2001 .